# Light Intensity Modulated Impedance Spectroscopy (LIMIS) in All-Solid-State Solar Cells at Open Circuit

Osbel Almora[1,2,3]*, Yicheng Zhao[1], Xiaoyan Du[1], Thomas Heumueller[1], Gebhard J. Matt[1], Germà Garcia-Belmonte[3] and Christoph J. Brabec[1]

[1]*Institute of Materials for Electronics and Energy Technology (i-MEET), Friedrich-Alexander-Universität Erlangen-Nürnberg, 91058 Erlangen, Germany;*

[2]*Erlangen Graduate School in Advanced Optical Technologies (SAOT), Friedrich-Alexander-Universität Erlangen-Nürnberg, 91052 Erlangen, Germany*

[3]*Institute of Advanced Materials (INAM), Universitat Jaume I, 12006 Castelló, Spain*

**osbel.almora@fau.de*

## Abstract

Potentiostatic impedance spectroscopy (IS) is a well stablished characterization technique for elucidating the electric resistivity and capacitive features of materials and devices. In the case of solar cells, by applying a small voltage perturbation the current signal is recorded and the recombination processes and defect distributions are among the typical outcomes in IS studies. In this work a photo-impedance approach, named "light intensity modulated impedance spectroscopy" (LIMIS), is first tested in all-solid-state photovoltaic cells by recording the individual photocurrent (IMPS) and photovoltage (IMVS) responsivity signals due to a small light perturbation at open-circuit (OC), and combining them: LIMIS=IMVS/IMPS. The experimental LIMIS spectra from silicon, organic, and perovskite solar cells are presented and compared with IS. An analysis of the equivalent circuit numerical models for total resistive and capacitive features is discussed. Our theoretical findings show a correction to the lifetimes evaluations by obtaining the total differential resistances and capacitances combining IS and LIMIS measurements. This correction addresses the discrepancies among different techniques, as shown with transient photovoltage. The experimental differences between IS and LIMIS (i) proves the unviability of the superposition principle, (ii) suggest a bias-dependent photo-current correction to the empirical Shockley equation of the steady-state current at different illumination intensities around OC and (iii) are proposed as a potential figure of merit for characterizing performance and stability of solar cells. In addition, new features are reported for the low-frequency capacitance of perovskite solar cells, measured by IS and LIMIS.

## 1. Introduction

Standard potentiostatic impedance spectroscopic (IS) is a well-known and stablished technique for the characterization of the resistive, capacitive and inductive features of materials and solar cells.[1, 2] In photovoltaic devices, one of the most common characterization routines is to probe the open-circuit (OC) condition under an steady-state illumination intensity and by applying a small voltage perturbation at different light intensities the IS spectra are measured and analyzed. In this way the recombination resistance $R_{rec}$, chemical capacitance $C_\mu$ and characteristic lifetimes $\tau$ are typically accessed.

With an alternative approach, the photo-sensitive samples have been earlier separately characterized by means of the intensity modulated photocurrent spectroscopy (IMPS)[3-14] and the intensity modulated photovoltage spectroscopy (IMVS).[9, 10, 15-17] Particularly, the IMPS has been recently gaining attention in the field of perovskite solar cells (PSCs), mainly exploring the short-circuit (SC) condition.[13, 14, 18-20] IMVS and IMPS individually characterize the current and voltage responsivities $\Psi_J$ and $\Psi_V$, respectively. Here a mere dimension analysis suggests that IMVS/IMPS has units of Ohms, like the impedance $Z$ from IS. Therefore, it may be interesting to analyze IS and this ratio, here-on called light intensity modulated impedance spectroscopy (LIMIS). Purposely, Song & Macdonald[21] first introduced and measured the concept on n-Si in KOH solution, validating the transfer function by Kramers-Kronig transformation. Also Halme[22] tackled the subject and measured IMVS/IMPS in dye sensitized solar cells, concluding the approximate equivalence with IS. More recently, Bertoluzzi & Bisquert[12] mentioned the concept but only analyzed separately IMVS,

IMPS and IS in water splitting systems. Simultaneously to this work, we have proposed an analytical model which shows the difference between LIMIS and IS to be proportional to the surface recombination velocity.[23]

In this article we further analyze this concept at OC and first present an experimental analysis of LIMIS silicon,[24] organic and perovskite solar cells.[25] The differences between IS and LIMIS spectra are introduced as a figure of merit for characterizing performance and degradation in solar cells. Our theoretical results suggest corrections to the concepts of differential resistance $R$ and capacitance $C$ for photosensitive samples under illumination. We show how by neglecting LIMIS the $R$ and $C$ can be over- and under-estimated, respectively, which ultimately corrects the assessment of charge carrier lifetimes. Our correction tackles the issue of the differences between experimental lifetime results from different techniques, as shown for transient photovoltage (TPV) profiles. The general equivalent circuit (EC) numerical model for the interpretation of the total $R$ and $C$ from photosensitive devices is also introduced, and the main differences in terms of EC fitting between IS and LIMIS are discussed. In addition, new capacitive features are reported for the low-frequency capacitance of PSCs. Finally, a bias-dependent photo-current correction is proposed to the empirical Shockley equation in order to conciliate the experimental observations and the theoretical deductions at OC for the steady-state current density-voltage $J - V$ curves at different light intensities.

In the following sections first IS and then LIMIS will be introduced in detail. The results are structured from the experimental reports to the theoretical deductions and analyses, and the conclusions are offered subsequently. Both theoretical and experimental results are significantly complemented with the online supporting information. Note the list of acronyms, symbols and abbreviations in Table S1 in order to facilitate the reading.

## 1.1. Potentiostatic impedance spectroscopy (IS) in solar cells

We may first consider a generic sample at steady-state voltage $\bar{V}$ where a current density $\bar{J}(\bar{V})$ is flowing. In a first approximation, every sample can be assumed as a resistor-capacitor $RC$ Voigt element with a characteristic time response constant $\tau = RC$, as in **Figure 1**a. Then a small potentiostatic perturbation $\tilde{V}(t) = |\tilde{V}|\exp[i\,\omega t]$ can be applied in alternating current ($ac$) mode, being $t$ the time, $\omega$ the angular frequency and $i$ the imaginary unit. The total voltage would be

$$V = \bar{V} + |\tilde{V}|\exp[i\,\omega t] \tag{1}$$

Upon perturbation, the current may evolve as

$$J = \bar{J} + \tilde{J}\exp[i\,\omega t] \tag{2}$$

where $\bar{J}$ may be the steady-state current $\bar{J}(\bar{V})$ and the phasor-related part $\tilde{J}$ may inform on the differential resistive and capacitive features of the sample. A typical sinusoidal $\tilde{V}(t)$ small perturbation is illustrated in **Figure 1**b, to which the current may be $\phi$ phase shifted, as in **Figure 1**c. Then we can write $\tilde{J} = |\tilde{J}|\exp[-i\phi]$ and the impedance can now be introduced as

$$Z(\omega) = \frac{\tilde{V}(t)}{\tilde{J}(t)} = \frac{|\tilde{V}|}{|\tilde{J}|}\exp[i\phi] \tag{3}$$

The $\phi$-dependence on frequency $f = \omega/2\pi$ creates an impedance spectrum, which is the study subject of the impedance spectroscopy (IS). Most typically presented as $Z(\omega) = Z'(\omega) + i\,Z''(\omega)$, the Nyquist plot representation is illustrated in **Figure 1**d.

There the characteristic semicircle from a linear $RC$ couple with single $\tau$ is shown. The real part $Z'$ carries the information on the differential resistance, and since $\phi \to 0$ when $\omega \to 0$ thus $Z \to Z'$ and the total differential resistance can be taken as the radius of the semicircle. On the other hand, the imaginary part $Z''$ informs on the capacitive features. Note that the $-Z''$ maximum ($\phi = \pi/4$ in **Figure 1**d) belongs to the characteristic angular frequency $\omega_\tau = \tau^{-1} = (R \cdot C)^{-1}$.

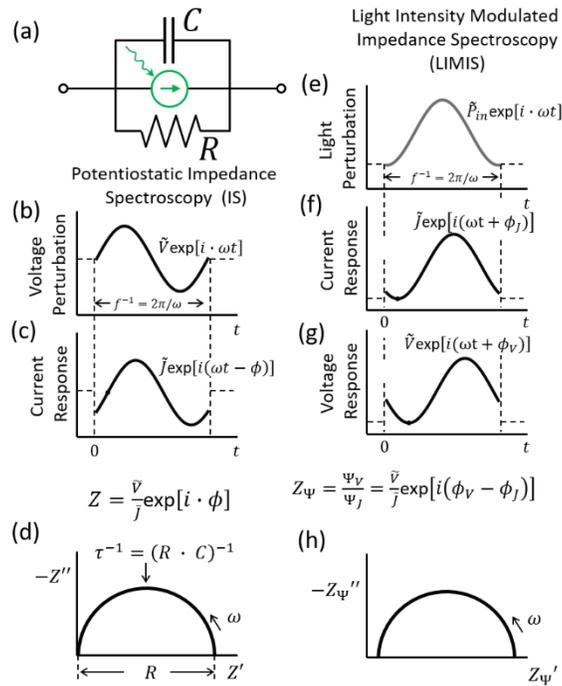

**Figure 1.** Schematized strategies for perturbation and characterization of electric responses from a solar cell as a (a) photo-sensitive simple RC Voigt circuit element: (b-d) IS and (e-h) LIMIS. For simplicity the modulus notation was avoided.

For a solar cell, the series resistance $R_{series}$ and shunt resistance $R_{sh}$ effects could be ideally neglected at forward bias, under illumination and around the open-circuit (OC) regime condition. In those circumstances the current density-voltage $J - V$ characteristic can be taken as the empirical approximation of the Shockley equation

$$J = J_s \left(\exp\left[\frac{q V}{m k_B T}\right] - 1\right) - J_{ph} \tag{4.a}$$

where $q$ is the elementary electric charge, $k_B T$ the thermal energy, $J_s$ the saturation current, $m$ the ideality factor, and $J_{ph}$ is the bias-independent photo-generated current, typically taken as the short-circuit current $J_{sc}$. Equation (4.a) characterizes the experimental measurement where the current is considered in steady state, i.e. direct current (*dc*) mode. Particularly, at open-circuit (OC) under forward *dc* biases larger than $5k_B T$ the expression (4.a) can be expressed in terms of the open circuit voltage $V_{oc}$ reciprocity

$$J_{ph} \cong J_s \exp\left[\frac{q V_{oc}}{m k_B T}\right] = \Psi_J P_{in} \tag{4.b}$$

where $\Psi_J = \Psi_{sc}$ is the photo-current responsivity at short-circuit that depends on the incident light spectrum, the absorption coefficient and the geometry of the absorbing materials, and $P_{in}$ is the incident light intensity in units of power density. Experimentally, under *ac* potentiostatic perturbation (IS measurement) the current signal results from evaluating (4.a) in (1), sampling the narrow region around the steady state condition. **Figure 2**a illustrates the particular case where OC is tested at constant illumination intensity.

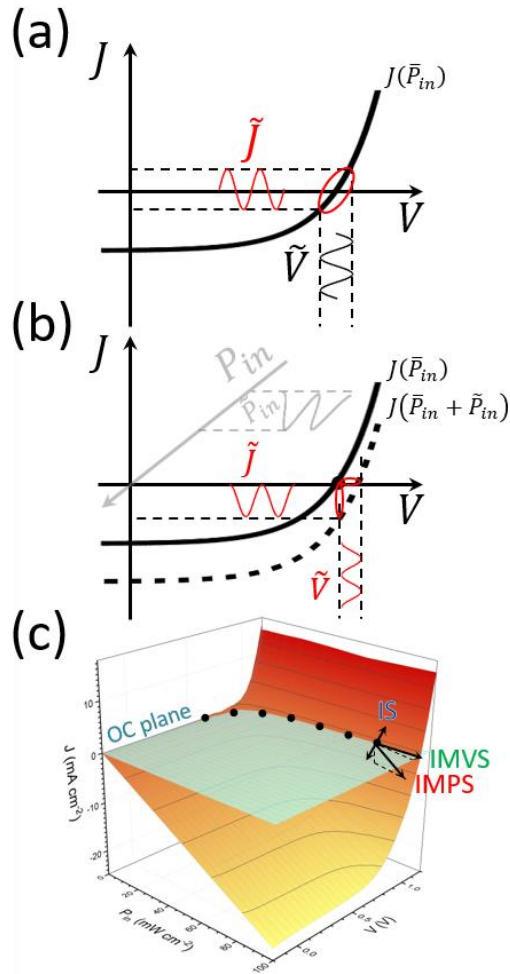

**Figure 2.** Schematized impedance characterization of a photovoltaic solar cell in 2D $J - V$ representations for (a) IS and (b) LIMIS. The thick dashed line in (b) is the projection of the perturbed $J(V, P_{in})$ curve in the 3D representation, as in (c) and (d). In (c) the three experimental measurement are illustrated at different points and in (d) there is a vector composition for the LIMIS, where IS, IMPS and IMVS characterize the same OC steady state.

### 1.2. Light intensity modulated impedance spectroscopy (LIMIS) in solar cells

Alternatively to the IS approach, in the case of photo-sensitive samples (see **Figure 1**a which included a photo-current source) the perturbation can be done by a light source. Then, a small perturbation $\widetilde{P}_{in}(t) = |\widetilde{P}_{in}|\exp[i\,\omega t]$ can be added to the given

$dc$ incident light power density $\overline{P}_{in}$, as in **Figure 1**e. The total incident light intensity in units of power density would be

$$P_{in} = \overline{P}_{in} + |\widetilde{P}_{in}|\exp[i\,\omega t] \tag{5}$$

Upon this perturbation, both current and voltage signals can be recorded. At a given $\overline{V}$, the current would be $\phi_J$ phase shifted, as in **Figure 1**f, and similarly to (2) $\widetilde{J} = |\widetilde{J}|\exp[i\phi_J]$. Hence a current responsivity transfer function can be defined as

$$\Psi_J(\omega) = \frac{\widetilde{J}}{\widetilde{P}_{in}} = \frac{|\widetilde{J}|}{|\widetilde{P}_{in}|}\exp[i\phi_J] \tag{6}$$

Likewise, at OC ($J = 0$) the photovoltage signal may be composed by the $dc$ open circuit voltage $\overline{V}_{oc}$ and the phasor related part as

$$V_{oc} = \overline{V}_{oc} + \widetilde{V}_{oc}\exp[i\,\omega t] \tag{7}$$

Then the photo-voltage signal may have a phase shift $\phi_V$ (see **Figure 1**g) and taking $\widetilde{V}_{oc} = |\widetilde{V}_{oc}|\exp[i\phi_V]$ thus a voltage responsivity transfer function is defined as

$$\Psi_V(\omega) = \frac{\widetilde{V}_{oc}}{\widetilde{P}_{in}} = \frac{|\widetilde{V}_{oc}|}{|\widetilde{P}_{in}|}\exp[i\phi_V] \tag{8}$$

Equations (6) and (8) define by themselves IMPS and IMVS, respectively. These techniques have been earlier introduced[3-6, 12, 16, 26] and there have been recent studies on photovoltaic solar cells.[10, 11, 13, 17, 27]

However, there are three main limiting factors when using individually IMPS or IMVS. First, the placing of a current (voltage) source for IMPS (IMVS) in the equivalent circuit (EC) is a particularly challenging task given that implies a direct impact in the distribution of currents and/or voltages which is not so straightforward in practice. This is an additional complication to the already debated selecting and justifying of using a given EC in IS, only with resistive and capacitive elements. Second, IMPS or IMVS cannot reproduce resistance or capacitance spectra, which limits it use in already stablished techniques like thermal admittance spectroscopy (TAS).[28] And last but not least, the validation of IMPS or IMVS with IS results is not so straightforward since each photo technique lacks one component or the other in terms of conductivity or field distribution.

Now, IMPS and IMVS can be combined to obtain the "light intensity modulated impedance spectroscopy" (LIMIS)

$$Z_\Psi(\omega) = \frac{\Psi_V}{\Psi_J} = \frac{|\tilde{V}_{oc}|}{|\tilde{J}|} \exp[i(\phi_V - \phi_J)] = |Z_\Psi|\exp[i\phi_\Psi] \qquad (9)$$

Advantageously, the experimental spectra from the photo-impedance of (9) do not need voltage/current sources in the EC-based numerical simulations, resulting a simpler and less ambiguous task. Also the spectroscopic representation of the resistive (see **Figure 1**h) and capacitive features can be obtained too, which allows future development of analogue light intensity modulated thermal admittance spectroscopy (LIMTAS). And furthermore, a LIMIS direct comparison with IS spectra may straightforwardly inform on generation/recombination features in solar cells.

**Figure 2**b presents a scheme on the LIMIS concept for typical photovoltaic cells -with $J - V$ characteristic as (4.a)- at OC under illumination. Despite the sampled *dc* condition is the same as for the IS (**Figure 2**a), LIMIS perturbates the steady-state in a third axis, corresponding to the incoming illumination power density $P_{in}$. As a result, the current and voltage signals spread individually, each one exclusively in the direction of its own axis. The thick dashed line in **Figure 2**b represent the 2-dimentional (2D) projection of the perturbated current, and **Figure 2**c shows the Equation (4.a) approximation for the family of 3-dimentional (3D) $J-V$ curves which are sampled when modulating incident light intensity. Also in **Figure 2**c the three separately examples of measurements are indicated: (i) IS in a *J-V* plane at a fixed $P_{in}$, (ii) IMPS in a $J-P_{in}$ plane at a fixed $V$ (short-circuit in the figure), and (iii) IMVS in the $V-P_{in}$ plane at open circuit, named OC surface.

Importantly, in the core of our focus is to provide a first approach to the difference between LIMIS and IS, its meaning and possible use. Accordingly, herein we define a normalized figure of merit called photo-impedance difference as

$$\Delta Z_\Psi = \frac{Z_\Psi - Z}{Z} \qquad (10)$$

where $Z$ and $Z_\Psi$ come after (3) and (9), respectively. Note that $\Delta Z_\Psi$ is zero when $Z_\Psi = Z$ and positive (negative) when the photo-impedance from LIMIS is larger (lower) than that from IS.

## 2. Results and discussion

### 2.1. Experimental LIMIS and IS spectra

A proper analysis between IS and LIMIS at OC requires to set the same steady-state *dc* illumination intensity. Subsequently IMVS can be measured directly at OC and for IS and IMPS the forward bias corresponding to the same $V_{oc}$ should be applied so the $J_{sc}$ is cancelled. For IMPS and IMVS, the exact set of sampled frequencies is an obvious requirement. Other external parameters like temperature, humidity (when reactivity issues) or even the wire connections should be controlled to be the same during the three measurements, so the characterized state is nearly the same.

The IS, IMPS and IMVS measurements were carried out with the Zahner Zennium Pro/PP211 impedance setup using its LSW-2 white LED light source. In all cases the sample holder included $N_2$ atmosphere.

Notably, ensuring the requirement of linear small perturbation is of upmost importance, mainly when measuring IMPS and IMVS to obtain LIMIS. In the case of IS, typically $\widetilde{V} < k_B T$ delivers accurate results, and for IMPS and IMPV keeping the *ac* perturbation below 10% of *dc* light intensity ($\widetilde{P}_{in} < \overline{P}_{in}/10$) provides a good empirical reference too. However, the latter rule can be not good enough in some cases, particularly for low *dc* illuminations approaching the *ac* experimental setup limit. Therefore, we use a significance parameter as described by Schiller and Kaus[29] and automatically implemented in the Zahner setup. The significance parameter goes from 0 to 1 and informs of "perfect linearity" if it equals unity. In practice, optimal results should be abode 0.98 and those below 0.95 should be discarded.

Five representative samples were experimentally characterized as summarized in Section S1.1: a silicon solar cell SiSC, an organic solar cell (OrgSC) and three perovskite solar cells (PSC1,2,3). The respective schemed structures, $J - V$ curves, external quantum efficiency (EQE) spectra and 500 hours degradation tests (for the PSCs) are in Figure S1. The performance parameters are in Table S2.

*The silicon device* constitutes the first reference due to the simplicity and robustness of its working principles. Its characterization is presented in Section S1.2: first the

IMPS and IMVS spectra at OC under different *dc* illumination intensities in Figure S2, and then LIMIS and IS spectra for the SiSC are shown in Figure S3. The current and voltage responsivities in Figure S2 illustrate the expected arc-like shapes in the Nyquist representation. In **Figure 3**a the low frequency limits from $\Psi_V$ and $\Psi_J$ are plotted. From IMVS the relation $\Psi_V = mk_B T q^{-1} P_{in}^{-1}$ with $m \approx 1.3$ is in agreement with theoretical predictions[16, 23] and the photocurrent-photovoltage trend in Figure S4a. On the other hand, from IMPS the light-intensity independency of $\Psi_J$ in almost all the range of measurement seems to fade only as $V_{oc}$ approaches the built-in voltage $V_{bi}$, illustrated in the Mott-Schottky plot of Figure S4b.

By applying the LIMIS definition (9) the photo-impedance spectra can be compared with the standard IS spectra, as in Figure S3 and **Figure 3**b. A right-shifted Nyquist plot is apparent, reporting $\Delta Z_\Psi > 0$ in the measured range. For the sake of clarity, this series-resistance-like right-shift in the real part of the LIMIS impedance is going to be referred in the next as $Z_s'$.

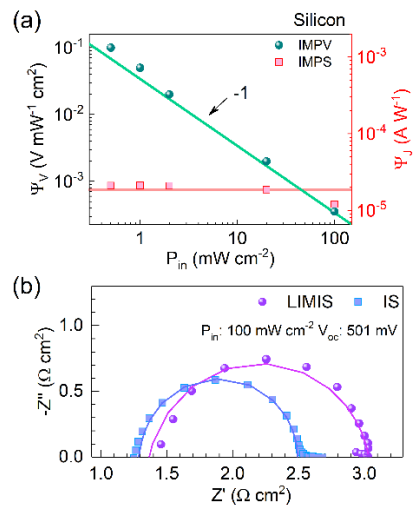

**Figure 3.** Silicon solar cell spectroscopic characterization: (a) low frequency limits of the voltage and current responsivities for different light intensities (see Figure

S2) and (b) representative impedance Nyquist plot (see Figure S3). Lines in (a) belong to fitting to $\Psi_V \propto P_{in}^{-1}$ and $\Psi_J \propto P_{in}^{0}$, and in (b) refers to the EC model discussed in Section 2.3 and **Figure 7**b.

*The organic device*, with structure ITO/ZnO/PM6:Y6/MoOx/Ag, was characterized as presented in Section S1.3 including the device fabrication description. Similarly, Figure S5 shows the IMPS and IMVS spectra and Figure S6 the comparison between IS and LIMIS.

**Figure 4**a presents the low frequency limits of the voltage and current responsivities spectra for the OrgSC. From the IMVS, $\Psi_V = mk_B T q^{-1} P_{in}^{-1}$ with $m \approx 1.8$ is again in agreement with theoretical predictions[16, 23] and the photocurrent-photovoltage trend in Figure S4c. Differently, from the IMPS $\Psi_J$ behaves more like $\Psi_J \propto P_{in}^{-1/3}$ in a low and medium range for the measured illumination intensities. Moreover, **Figure 4**b illustrate one of the Nyquist plots showing the similar arcs of the two techniques, also with the right-shifting trend for the LIMIS spectrum. More interestingly here it is that the apparent series resistance $Z_s'$ shows a negative arc in the Nyquist representation (empty dots in **Figure 4**b). This is an important feature whose understanding, while beyond the scope of this paper, should be attended in future works.

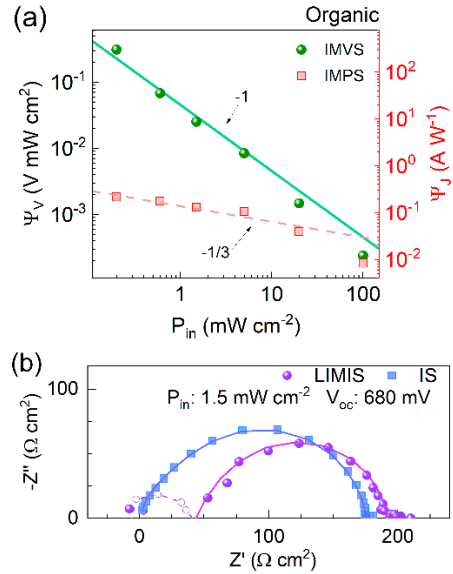

**Figure 4.** Organic solar cell spectroscopic characterization: (a) low frequency limits of the voltage and current responsivities for different light intensities (see Figure S5) and (b) representative impedance Nyquist plot (see Figure S6). Lines in (a) belong to fitting to $\Psi_V \propto P_{in}^{-1}$ and $\Psi_J \propto P_{in}^{0}$ (dashed line to $\Psi_J \propto P_{in}^{-1/3}$), and in (b) refers to the EC model discussed in Section 2.3 and **Figure 7**b. LIMIS empty dots in (b) mean negative values.

*The perovskite solar cells* under study all shared similar n-i-p structures, with variations in the absorber and the electron selective contact layers. Labeled as PSC1, we first discuss the spectroscopic characterization (see Section S1.4) of the more efficient and stable sample (see Section S1.1), with structure ITO/SnO$_2$/PMMA(PCBM)/Cs$_{0.05}$MA$_{0.1}$FA$_{0.85}$Pb(I$_{0.85}$Br$_{0.15}$)$_3$/PDCBT/Ta-WO$_x$/Au. Similarly, the characterizations of PSC2 and PSC3, with structures ITO/SnO$_2$/Cs$_{0.05}$MA$_{0.1}$FA$_{0.85}$Pb(I$_{0.85}$Br$_{0.15}$)$_3$/PDCBT/Ta-WO$_x$/Au and ITO/SnO$_2$/Cs$_{0.15}$FA$_{0.85}$PbI$_3$/PDCBT/Ta-WO$_x$/Au, are presented in sections S1.5 and S1.6, respectively.

The IMPS and IMVS spectra for PSC1 are presented in Figure S8, evidencing already a more complex response including two arcs in the Nyquist plots. Regarding the low frequency limits of the voltage and current responsivities from PSC1, in **Figure 5**a, the IMVS similarly gives $\Psi_V = mk_BTq^{-1}P_{in}^{-1}$ with now $m \approx 1.5$ following theory[16, 23] and agreeing previous reports on ideality factors from mixed cation PSCs.[30, 31] Distinctly, the IMPS reports a situation somehow in the middle between constant $\Psi_J$ at lower light intensities and $\Psi_J \propto P_{in}^{-1/3}$ at higher illuminations. The latter resembles the behavior of the OrgSC, probably related with the intrinsic absorber nature of both.

Applying LIMIS definition (9) allows to compare it with the IS spectra, as in Figure S9. PSC1 brings a new feature to the impedance spectra by reporting a clear three RC constants, i.e., three arcs in the Nyquist plots and three steps in the capacitance Bode plots. Importantly, as illustrated in **Figure 5**b, the high frequency region of the spectra ($f$>1kHz) from LIMIS delivers negative values in the Nyquist plot (empty dots) and a consequent negative capacitance in the Bode plots of Figure S9. Hence, the expected high frequency arcs (plateau) of the LIMIS impedance (capacitance) spectra from PSCs are not so and instead suggest a higher complexity in terms of EC elements. The high frequency region from IS reproduces earlier described features.[32, 33] On the other hand, at low frequencies ($f$<1kHz) LIMIS seems to reproduce very well the IS spectra, in both the impedance Nyquist plot (2 arcs in **Figure 5**b) and the capacitance Bode plot (2 steps in **Figure 5**c).

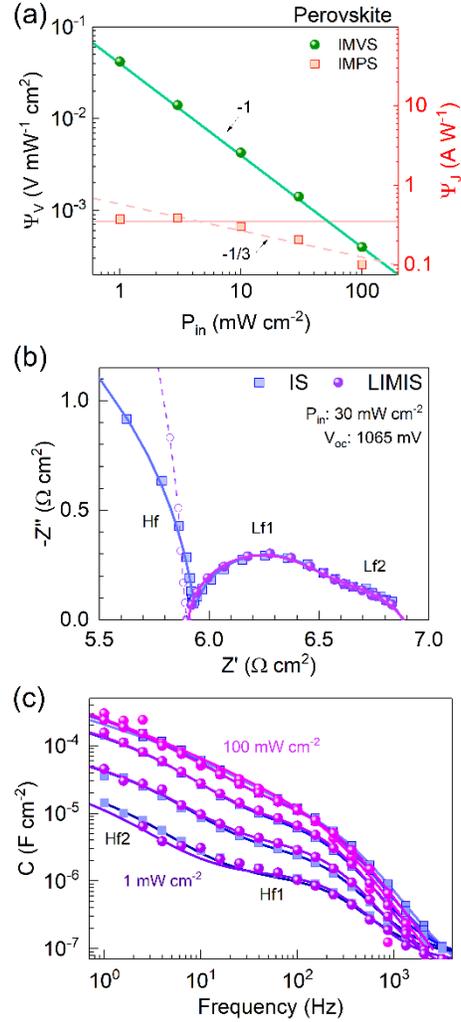

**Figure 5.** Perovskite solar cell (PSC1) spectroscopic characterization: (a) low frequency limits of the voltage and current responsivities for different light intensities (see Figure S2) and (b) representative impedance Nyquist plot (see Figure S3). Lines in (a) belong to fitting to $\Psi_V \propto P_{in}^{-1}$ and $\Psi_J \propto P_{in}^{0}$ (dashed line to $\Psi_J \propto P_{in}^{-1/3}$), and in (b) refers to the EC model discussed in Section 2.3 and **Figure 7**c. LIMIS empty dots in (b) mean negative values.

Moreover, one final interesting correlation is remarked regarding the comparison between LIMIS and IS in terms of $\Delta Z_\Psi$. By taking the low frequency limit we can express it as $\Delta Z'_\Psi = (Z_T' - R_T)/R_T$ where $Z_T'$ and $R_T$ come from LIMIS and IS respectively. This is displayed in **Figure 6** for the set of studied devices. The general trend shows first a decrease as light intensity is augmented until a few tens of mW·cm$^{-2}$. In this range a rough approximation would say that $Z_\Psi \propto Z(1 + P_{in}^{-2})$. Towards 1 sun illumination intensity, the photo-impedance from LIMIS seems to exceed the impedance from IS as light intensity grows. In this latter range we could speculate that $Z_\Psi \propto Z(1 + P_{in}^{2})$. Interestingly, in the region between the two regimes, some negative values are reported, indicating that the photo-resistance from LIMIS is lower that the total resistance from IS. This only occurs for the OrgSC, PSC2 and PSC3. These are actually the devices with more performance issues: the OrgSC presents "S" shape above OC and the PSC2 and PSC3, besides the lower PCE, and $V_{oc}$, showed lower stability too (see Figure S1). These correlations are also a matter of further analyses, but these preliminary observations suggest that the higher $\Delta Z'_\Psi$ the best, and that negative values of $\Delta Z'_\Psi$ indicate performance and/or degradation issues in solar cells.

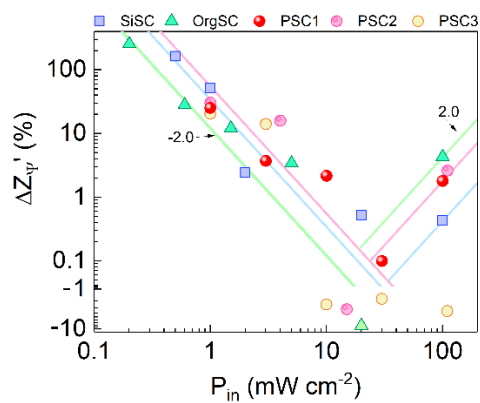

**Figure 6.** $\Delta Z'_\Psi$ as a figure of merit for checking performance and*or degradation issues: normalized real difference between $Z'_T$ and $R_T$, from LIMIS and IS respectively, as a function of illumination intensity for the different studied devices. Only the cells with low performance or degradation issues show negative $\Delta Z'_\Psi$.

## 2.2. The differential approach to resistance and capacitance: correcting lifetimes

The derivative of the scalar current $J(V, P_{in})$ is measured in different directions $\hat{v}_1$, $\hat{v}_2$ and $\hat{v}_3$ by IS, IMPS and IMVS, respectively. Thus, we can calculate them by using the concept of directional derivative and the directions in the OC surface ($J = 0$). For IS the derivative is found in the direction of $\widetilde{V}$, so $\hat{v}_1 = (1,0)$ and for IMPS, in the direction of $\widetilde{P}_{in}$, so $\hat{v}_2 = (0,1)$. These are the well-known partial derivatives in the axes directions. However, IMVS is not a partial derivative of $J(V, P_{in})$ but $V_{oc}$. Hence we may redefine it as the directional derivative of $J(V, P_{in})$ in the direction $\hat{v}_3$ contained in the interception between the OC surface and $J(V, P_{in})$ (see **Figure 2**c). Consequently, we can now express the transfer functions of IS (3), IMPS (6), IMVS (8) and thus LIMIS (9) as derivatives at $\overline{V} = \overline{V}_{oc}$, respectively as

$$Z = \left(\vec{\nabla}_{\hat{v}_1} J\right)^{-1} = \left(\vec{\nabla} J \cdot (1,0)\right)^{-1} = \left(\frac{\partial J}{\partial V}\right)^{-1} \tag{11.a}$$

$$\Psi_J = \vec{\nabla}_{\hat{v}_2} J = \vec{\nabla} J \cdot (0,1) = \frac{\partial J}{\partial P_{in}} \tag{11.b}$$

$$\Psi_V = \vec{\nabla}_{\hat{v}_3} J = \vec{\nabla} J \cdot (v_{30}, v_{31}) = \frac{\partial V_{oc}}{\partial P_{in}} \tag{11.c}$$

$$Z_\Psi = \frac{\partial V}{\partial P_{in}} \left(\frac{\partial J}{\partial P_{in}}\right)^{-1} \tag{11.d}$$

Purposely, we are here interested in two main physical quantities: the total differential resistance unit area

$$R = \left(\frac{dJ}{dV}\right)^{-1} \tag{12}$$

and the total differential capacitance per unit area

$$C = \frac{dQ}{dV} \quad (13)$$

where $Q$ is the charge density. Equations (12) and (13) are total differentials that can be approached to the partial derivatives from the potentiostatic IS following (11.a) as

$$R_{IS}(\omega) \approx \left(\frac{\partial J}{\partial V}\right)^{-1} = \text{Re}[Z(\omega)] \quad (14.a)$$

$$C_{IS}(\omega) \approx \frac{\partial Q}{\partial V} = \text{Re}\left[\frac{1}{i\,\omega\,Z(\omega)}\right] \quad (14.b)$$

where $Z(\omega)$ is that of (3) and $\partial Q \propto \partial J/\omega$ at each frequency. Definition (14) is the full form of (12) and (13) in dark measurements and for non-photosensitive samples.

An interesting exercise is to apply (14.a) to (4.a) in order to obtain the typically called $dc$ resistance

$$R_{dc} = R_{th} \exp\left[-\frac{q\,V}{m\,k_B T}\right] \quad (15.a)$$

where $R_{th} = m\,k_B T/J_s q$ is the thermal recombination resistance. In practice, IS resolves the different $\omega$-components in the resistive response from a sample, but the total resistance should resemble $R_{dc}$ and converge in the appropriate low frequency limit. Note that at OC $V$ should be substituted by $V_{oc}$ in (15.a) and $R_{dc}(V) = R_{dc}(V_{oc})$ only if $J_{sc}$ does not depend on bias.

We can also apply (11.b-d) to the empirical approximation of the Shockley equation (4) resulting the analogue $dc$ parameters

$$\Psi_{J,dc} = -\Psi_J \tag{15.b}$$

$$\Psi_{V,dc} = \frac{m \, k_B T}{q \, P_{in}} \tag{15.c}$$

$$R_{\Psi,dc} = \frac{m \, k_B T}{q \, \Psi_J P_{in}} = R_{th} \exp\left[-\frac{q \, V}{m \, k_B T}\right] \tag{15.c}$$

where $\Psi_J = J_{sc}/P_{in}$ has the same meaning as in (4.b). Note that in the assumption of bias-independent $\Psi_J$, and in agreement with (4.b), $\Psi_J P_{in} = J_s \exp[qV/m \, k_B T]$ and thus $R_{dc} = R_{\Psi,dc}$. This equality expresses the dark/light superposition rule, following the reciprocity theorem of charge collection,[34] which implies that the $J-V$ curves under illumination are the same as the dark one, only current shifted an amount $-J_{sc}$.

Now, similarly to (14) for IS, the IMPS and IMVS respectively explore partial derivatives as in (11.b,c). This is illustrated in **Figure 2**c and left side of **Figure 2**d. Subsequently, since (4) is not light independent, the definition (12) can be better approached as

$$R = \left(\frac{\partial J}{\partial V} + \frac{\partial J}{\partial P_{in}} \frac{\partial P_{in}}{\partial V}\right)^{-1} \tag{16.a}$$

$$R(\omega) = \left(\frac{1}{\mathrm{Re}[Z(\omega)]} + \frac{1}{\mathrm{Re}[Z_\Psi(\omega)]}\right)^{-1} = \left(\frac{1}{R_{IS}(\omega)} + \frac{1}{R_\Psi(\omega)}\right)^{-1} \tag{16.b}$$

From (16), note that the predominant term will be the lower of the resistances $R_{IS}$ and $R_\Psi$, from IS and LIMIS respectively. Also, from the *dc* approximation, if $R_{IS} \sim R_\Psi$ then $R$ results around a half of that typically estimated from IS. This result is related with the typical photoconductivity enhancement in photovoltaic devices when comparing dark and light behavior by IS or even *dc* measurements. $R$ is a measure of how much current ($dQ/dt$) changes per unit change of voltage, and (16) express the light dependency of that ratio.

On the other hand, in the case of the capacitance, it makes sense to think that, some extra charge is stored in the device under illumination, different to what would be expected from the dark regime, even considering chemical capacitance effects. Accordingly, a better estimation of the differential capacitance in photosensitive samples would be

$$C = \frac{\partial Q}{\partial V} + \frac{\partial Q}{\partial P_{in}} \frac{\partial P_{in}}{\partial V} \qquad (17.a)$$

$$C(\omega) = \text{Re}\left[\frac{1}{i\,\omega\,Z(\omega)} + \frac{1}{i\,\omega\,Z_\Psi(\omega)}\right] = C_{IS}(\omega) + C_\Psi(\omega) \qquad (17.b)$$

Similarly, from (17), note that the predominant term will be the larger of the capacitances $C_{IS}$ and $C_\Psi$, from IS and LIMIS respectively. Specifically, if $C_{IS} \sim C_\Psi$ then $C$ results twice that typically estimated from IS. As in the case of resistance, light charges the capacitor in addition to how the bias does it, hence it makes sense that some extra charge is stored. Accordingly, it is of crucial importance to evaluate the degree of overestimation (underestimation) of the differential resistance (capacitance) by only considering IS measurements.

Importantly, if the superposition rule (14) is valid, then $Z = Z_\Psi$ makes $R = R_{IS}/2$ and $C = 2C_{IS}$. Accordingly, the LIMIS measurements would not modify the corresponding characteristic response times $\tau = RC = R_{IS}C_{IS}$. For instance, this would be the case where the characteristic lifetimes from IS spectra coincide with some other techniques like TPV, as earlier reported.[35] However, as showed in the previous section, we found $Z \sim Z_\Psi$ which may deliver a corrected lifetime including all the carrier contributions due to bias and light dependencies. Note that this result does not conflicts the reciprocity theorem of charge collection,[34] which states the equivalence between the currents due to photo-generation at a point surrounded by no charge and the injection of the same charge to the surrounding of the same point, if all the rest of boundary conditions in the space are kept the identical. Our findings of $Z \neq Z_\Psi$ just reflect how photo-generation modifies the boundary conditions with respect to dark recombination currents due to the injection of carriers.

### 2.3. Numeric approach: the equivalent circuits

After introducing LIMIS in Section 1.2, the total differential resistance and capacitance from photosensitive samples was corrected in Section 2.2, resulting as in equations (16) and (17). Differently to the that suggested by the derivatives of the *dc* empirical Shockley equation (15), our recent analytical analysis[23] suggested that the impedances from IS and LIMIS should differ. Accordingly, the accurate estimation of $R$ and $C$ may include the measurement of LIMIS. In particular, the incorporation of both concepts can be represented in an equivalent circuit (EC) as in **Figure 7**a, where the impedance $Z$ from IS and the photo-impedance $Z_\Psi$ from LIMIS are connected in parallel among them, excluding the non-photosensitive contributions from the ohmic series resistances $R_{series}$. Also in **Figure 7**a the simplest EC including a couple of Voigt elements in parallel is illustrated, in agreement with differential definitions (16) and (17).

IS and LIMIS are measured separately, and in the next section the experimental measurement will be presented and discussed. **Figure 7**b,c display the ECs used for the numerical fitting indistinctively for IS and LIMIS. These are well-known widely used ECs for characterizing solar cells.[2, 30, 36] Note that even in cases where both techniques were fitted with the same EC, that does not mean that every element share the same physical meaning. In any case, from **Figure 7**b,c the total resistances either from IS or LIMIS are taken from the series connection as $R_T = R_{\text{Hf}} + R_{\text{Lf}}$ and the total capacitance, in parallel, as $C_T = C_{Hf} + C_{Lf}$. In most of the cases $C_{Hf}$ will be the geometrical capacitance of the sample $C_g$, or the depletion layer capacitance $C_{dl}$. From these parameters, the characteristic times $\tau_{Hf} = C_{Hf} R_T$ and $\tau_{Lf} = C_{Lf} R_{Lf}$ can be obtained and studied.

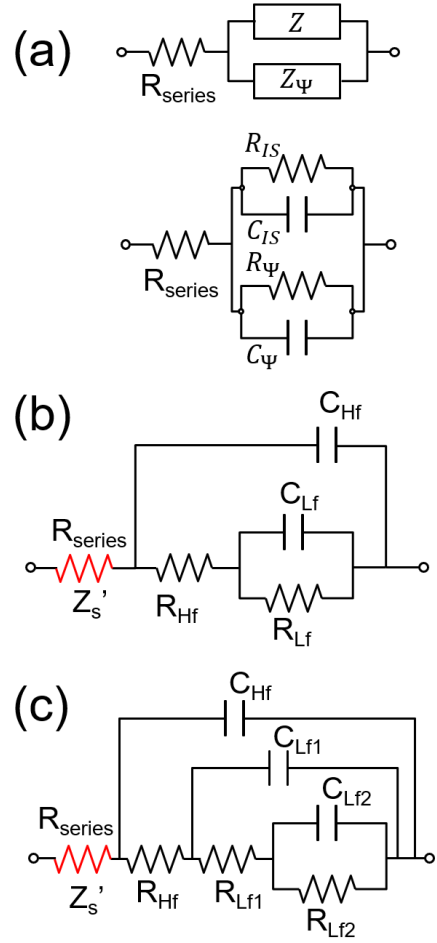

**Figure 7.** Equivalent circuits for (a) the concept of total differential contributions from IS and LIMIS to resistance and capacitance and (b, c) the used equivalent circuits during the numerical simulation of IS and LIMIS spectra. $R_{series}$ is a series resistor, $R_{IS}$ and $R_\Psi$ are the total resistances measured by IS and LIMIS, $R_{Hf}$ and $R_{Lf}$ are high and low frequencies resistors, $C_{IS}$ and $C_\Psi$ are the total resistances measured by IS and LIMIS, and $C_{Hf}$ and $C_{Lf}$ are high and low frequencies capacitors, respectively.

*For the silicon solar cell*, the IS and LIMIS spectra were numerically simulated to the EC model of **Figure 7**b as presented with solid lines in Figure S3 and **Figure 3**b. The total $C$-coupled resistances $R_T = R_{Hf} + R_{Lf}$ are shown in **Figure 8**a as a function of the $V_{oc}$. $R_T$ follows an exponential law like $R_{dc}$ (15.a) with $m \approx 1.2$, being $R_{th}$ approximately a 20% larger for LIMIS than that from IS, i.e., wider arcs as in **Figure 3**b. Accordingly, from (16): $R \approx 0.6 \cdot R_{IS}$ under illumination.

In addition, the right-shifting $Z_s'$ (see **Figure 3**b) also follows an exponential decrease as (15.a), but with $m \approx 2$, as in **Figure 8**a. This is an extra impedance contribution, different than that of the ohmic $R_{series}$ (nearly constant in **Figure 8**a) which may be detailed studied in the future. Here it is important to note that the high frequency part of the LIMIS spectra is particularly difficult to fit due to the lower linearity of the signal, as expressed in the significance spectra of Figure S3.

The capacitance bode plots are shown in Figure S3 with the respective simulations to **Figure 7**b EC model. The total capacitance $C_T = C_{Hf} + C_{Lf}$ from the fittings is plotted in **Figure 8**b showing an exponential increase possibly due to diffusion capacitance.[32,37] In this case the $C_T$ from LIMIs is nearly half of that from IS, so from (17): $C \approx 1.5 \cdot C_{IS}$ under illumination.

With the information of $R$ and $C$, the characteristic times $\tau = RC$ can be accessed, as presented in **Figure 8**c. Note that from (16) and (17) the total time response is actually a 90% of that calculated for IS. **Figure 8**c also present the transient photovoltage (TPV) lifetimes $\tau$ which nearly coincides with IS and LIMIS a lower light intensities (below ~3 mW·cm$^{-2}$). The TPV measurements were performed with a self-made setup (see Section S1.7 for details) in order to contrast the results from the characteristic time constants. As the light intensity is augmented, the TPV signal does not decay exponentially anymore (see Figure S17) and the IS and LIMIS provide a better estimation of characteristic lifetimes. In addition, The LIMIS seems to inform on faster characteristic times ($\tau_{Hf}$), possibly related with charge extraction processes, i.e., not as slow as the recombination lifetime.

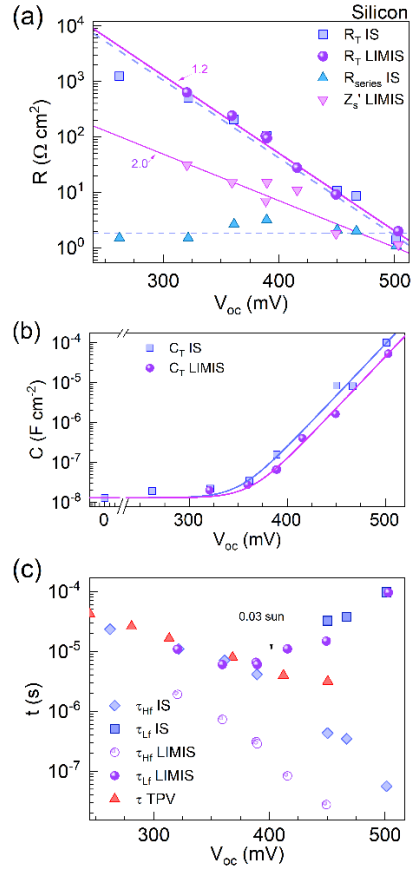

**Figure 8.** Silicon solar cell numerical simulation results: (a) resistance, (b) capacitance and (c) characteristic times from LIMIS and IS (EC model in **Figure 7**b) and TPV (mono-exponential decay model). The lines in (a) are the fittings to (15.a) with $m$ as indicated with arrows.

*For the organic solar cell*, **Figure 9**a shows $R_T$, $Z_s'$ and $R_{series}$ as a function of $V_{oc}$ for the IS and LIMIS spectra, as well as the numerical simulations to the EC model of **Figure 7**b, as presented with solid lines (see Figure S6). The OrgSC displays a more evident trend as $Z_s' \propto \exp[-qV_{oc}/2\,k_B T]$ and $R_T$ also behaves like $R_{dc}$ (15.a) but with $m \approx 1.8$ and $R_{th}$ approximately a 8% larger for LIMIS than that from IS, which gives from (16): $R \approx 0.51 \cdot R_{IS}$ under illumination.

Capacitance spectra are also displayed in Figure S6, and the total capacitance of the OrgSC from the fittings is presented in **Figure 9**b, which is basically $C_{Lf}$, significantly higher and exponentially increasing in comparison with the constant geometrical capacitance $C_{Hf} = C_g$. In this case LIMIS presents a 64% higher capacitance with respect to LIMIS, so from (16): $C \approx 2.64 \cdot C_{IS}$. Accordingly, the actual total characteristic times may be 1.35 times bigger than they are from IS, which is nearly $\tau_{Lf}$ for LIMIS. This result approaches the lifetimes from TPV below $\sim 10$ mW·cm$^{-2}$ (see decays in Figure S10) and the characteristic times from IS and LIMIS, as presented in **Figure 4**c. In that figure it is also evident how LIMIS and IS characteristic times are similar for the OrgSC, unlike the SiCS.

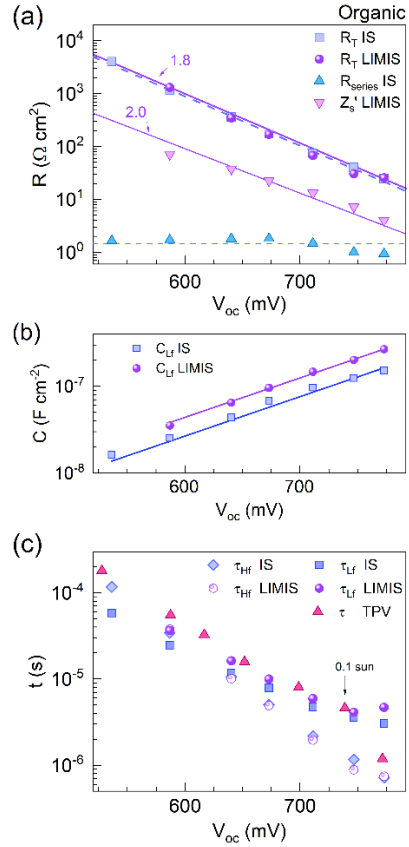

**Figure 9.** Organic solar cell numerical simulation results: (a) resistance, (b) capacitance and (c) characteristic times from LIMIS and IS (EC model in **Figure 7**b) and TPV (mono-exponential decay model). The lines in (a) are the fittings to (15.a) with $m$ as indicated with arrows.

*The perovskite solar cells* spectra were simulated with extra resistive and capacitive parameters, now using the EC model of **Figure 7**c. Note that such EC model does not include inductive elements, as earlier needed[30] in devices with similar mixed perovskite absorber but $TiO_2$ and spiro-OMeTAD as selective contacts. This suggest that the inductive behavior is an electrode-related issue.

The $R_T$ and $R_{series}$ from PSC1 IS spectra is summarized in **Figure 10**a. Note that $R_T \approx R_{Hf}$, meaning that the contributions to resistance from the $R_{Lf1}$ and $R_{Lf2}$ are much lower. But in the case of LIMIS in PSCs, the $R_{Hf}$ is mostly replaced by the series-resistance-like parameter $Z'_s$ and the total $C$-coupled resistances only include low frequency contributions $R_T \approx R_{Lf1} + R_{Lf1}$. Thus, what makes sense in PSCs is to compare $R_T$ from IS vs. $Z'_T = R_T + Z'_s$ from LIMIS. The resistive fitting parameters are summarized in **Figure 10**a showing $R_T$ and $Z'_T$ proportional to $\exp[-qV_{oc}/1.7\ k_BT]$, $Z_s' \propto \exp[-qV_{oc}/2.5\ k_BT]$ and $R_{series}$ stepped constant as a function of $V_{oc}$. Analogously, $Z'_T$ from LIMIS is around 15% larger than $R_T$ from IS, so from (16): $R \approx 0.53 \cdot R_{IS}$ under illumination.

More interestingly are the capacitive features and the resulting time constants. In **Figure 10**b, we show low frequency capacitances $C_{Hf1}$ and $C_{Hf2}$ as resulted from the numerical fitting to the EC model of **Figure 7**c. The trend $C_{Hf1} \propto \exp[qV_{oc}/1.5k_BT]$ has been earlier reported as a distinctive feature in mixed cations perovskite based solar cells with $TiO_2$ and spiro-OMeTAD as selective contacts.[30, 32] However, the even higher and saturating-like $C_{Hf2} \propto \exp[qV_{oc}/5k_BT]$ is a new finding. $C_{Hf2}$ may be connected to the modification of interface contact with the $SnO_2$ and/or the PDCBT and the types of cations composing the absorber layer. This adds extra elements to the already anomalous capacitive response of PSCs, closely connected with the $J-V$ curve hysteretic behaviors.[2, 38] Importantly, LIMIS and IS both nearly reproduce the same capacitances, which reinforce the idea of interconnected ionic-electronic nature of these slower mechanisms.

The characteristic response times are summarized in **Figure 10**c. The high frequency times follow the resistance trend and even approximately agree with the TPV lifetimes (see decays in Figure S14). The low frequency times behave slightly constant and decreasing, $\tau_{Hf1}$ and $\tau_{Hf1}$ respectively, suggesting an eventual convergence around miliseconds.

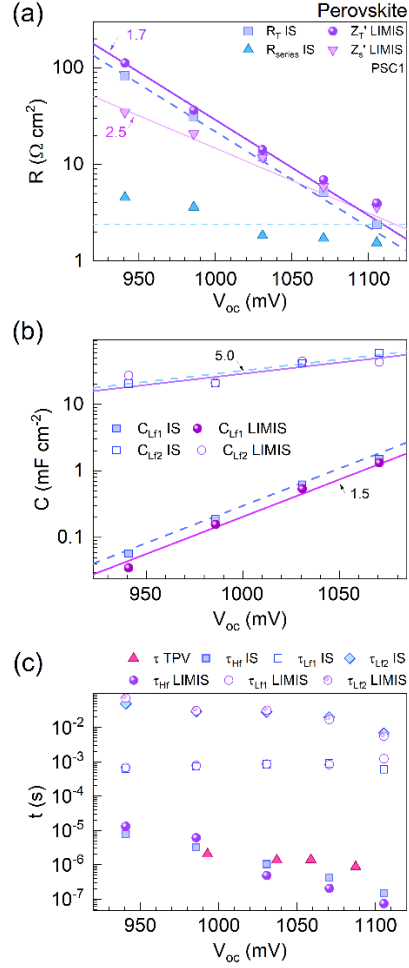

**Figure 10.** Perovskite solar cell (PSC1) numerical simulation results: (a) resistance, (b) capacitance and (c) characteristic times from LIMIS and IS (EC model in **Figure 7**c) and TPV (mono-exponential decay model). The lines in (a) are the fittings to the $R_{dc}$ behavior with m as indicated with arrows. The lines in (a) are the fittings to (15.a) with $m$ as indicated with arrows.

Two more perovskite solar cells PSC2 and PSC3 were analyzed as summarized in Sections S1.5 and S1.6, respectively, with nearly similar trends to PSC1. Nevertheless, regarding the low frequency capacitance, by eliminating the PMMA/PCBM cover towards the SnO₂ in PSC2, we obtain almost totally saturated $C_{Hf} = C_{Hf2}$ and discrepancies between LIMIS and IS (see Figure S13b), converging as light intensity increases. On the other hand, typical $C_{Hf} = C_{Hf1} \propto \exp[qV_{oc}/1.5k_BT]$ is again obtained if, with the same electrodes as PSC2, the methylamonium and Bromide compositions are neglected, as in PSC3 (see Figure S16b). The detailed analysis of these features, while only reported here, should be attended in future studies.

### 2.4. Bias-dependent photocurrent correction to the empirical Shockley equation around open circuit

From the previous section it was stated how the LIMIS spectra, despite resembling the IS shapes, are not the same as the IS spectra. This result from the spectroscopic *ac* characterization is also in agreement with the *dc* response in Section S1.8. In Figure S18 the experimental *J-V* curves from three of the studied samples (SiSC, OrgSC and PSC1) are presented as a function of the illumination intensity, forming current three-dimensional (3D) surfaces. The corresponding short-circuit currents are displayed in Figure S19 confirming the well-known relation $J_{sc} \propto P_{in}$ at $V = 0$ as (4.b).

From the experimental data in Figure S19 we can numerically find the pair $(V, P_{in})$ for the current roots (OC) and calculate the numerical derivatives for IS, IMPS and IMVS as (11). The code for that calculus is in Table S2 and the results are shown in **Figure 11** comparing IS and LIMIS as *dc* resistances. For the SiSC, IS and LIMIS coincide only at the highest illumination intensities, where $R_{dc} \propto P_{in}^{-1}$ agrees with **Figure 3a** suggesting light independent $\Psi_J$ and $P_{in} = J_s \exp[qV/mk_BT]$ from the empirical Shockley equation (4). For samples OrgSC and PSC1(and SiSC at lower $P_{in}$) the $R_{dc}$ is more evidently different from IS and LIMIS. Particularly, the "S" shape of OrgSC above OC creates a remarkable difference between IS and LIMIS. Interestingly, different trends $R_{dc} \propto P_{in}^{-b}$ are found depending on the sample and the illumination intensity range.

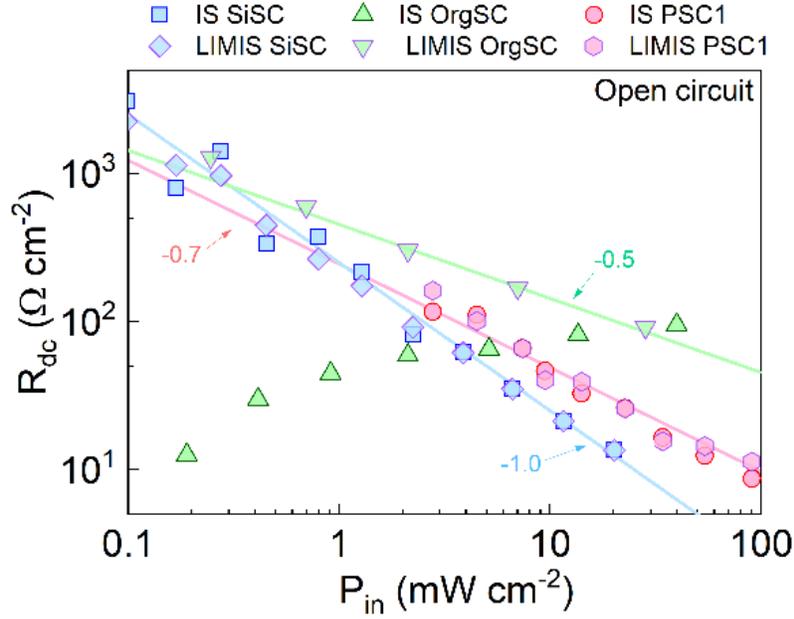

**Figure 11.** Numerically calculated *dc* resistances at OC using the differential definitions of IS and LIMIS on experimental $J-V$ curves at different light intensities. The experimental data is plotted in Figure S21 and the calculation code in Table S2.

The above experimental observations contradict the formulation of the well-known empirical Shockley equation (4) when applying the LIMIS definition (9) as in (15). Note first that we already showed $\Psi_J$ decreasing with light intensity at OC (see **Figure 4**a, **Figure 5**a, Figure S11d and Figure S14d), but at SC it still agrees with (4), as in Figure S19. Accordingly, a correction in (4.a) may be introduced justifying to apply the IMPS differential definition (11.b) at OC to obtain a decrease of $\Psi_J$ as $P_{in}$ increase. This correction makes sense only if a bias-dependency $\Psi_J(V)$ is included.

The deviations from the superposition principle in a form of bias-dependent photocurrent are well reported issues in silicon,[39, 40] thin film[41-44] and organic[45-47] solar cells. Another related subject is the *ac/dc* photo-shunting[48, 49] at SC, also reported for PSCs.[2, 32, 50] These are still open problems, which have been approached in several ways. In practice, under illumination both the photogenerated as well as recombination and drift-diffusion current components are modified by the field profile. Thus, only by numeric simulations the actual collection efficiency can be evaluated. Nevertheless, it is customary to still neglect changes in the dark diode term in (4) and group all the corrections to the model in the $J_{ph}$ term, which can be experimentally accessed from the difference between dark and illuminated $J - V$ curves. In this direction, an empirical formulation would be

$$J_{ph} \cong J_{sc} \left( \varsigma + \frac{\varsigma - 1}{1 + \exp\left[\frac{q(V - V_\varsigma)}{m_\Psi k_B T}\right]} \right) \tag{18}$$

with $J_{sc} = P_{in}\Psi_{sc}$, the collection fraction $\varsigma$, the collection threshold voltage $V_\varsigma$ and $m_\Psi > m$ is the photocurrent ideality factor. Here $V_\varsigma$ indicates the critical bias above which the $J_{sc}$ loses are more than half. Since the flat-band condition is particularly detrimental for the charge extraction, it makes sense to approach it to the built-in voltage $V_\varsigma \sim V_{bi}$. The parameter $\varsigma$ signifies how much photocurrent holds upon bias increasing: $\varsigma = 1$ means no loses, $\varsigma = 0$ indicate loss of entire $J_{sc}$ and $\varsigma < 0$ implies a crossing between light and dark $J - V$ curves (see Figure S20a). The step-like expression (18) successfully describes most of the experimental behaviours, but it precludes finding an analytical expression for the $V_{oc}$ analogue to (4.b) to calculate derivatives as (11). However, our evidence and general focus is only around the OC regime, thus, what makes sense is to approximate (18)

Particularly from our observations around OC we can empirically approximate

$$\Psi_J = \Psi_{oc}\exp\left[-\frac{qV}{m_\Psi k_B T}\right] \quad (19)$$

where $\Psi_{oc} \approx \Psi_{sc}\exp[qV_\varsigma/m_\Psi k_B T]$ is the current permittivity at OC. With this assumption, the corrected empirical Shockley equation around OC should be reformulated as

$$J \cong J_s\left(\exp\left[\frac{qV}{m\,k_B T}\right] - 1\right) - P_{in}\Psi_{oc}\exp\left[-\frac{qV}{m_\Psi k_B T}\right] \quad (20)$$

Equation (20) successfully reproduces the photocurrent around $V_{oc}$, as illustrated in Figure S20b,c. Hence, we can rewrite (15) for OC condition (see deductions in Section S2.2) as

$$R_{dc} = R_{th}\exp\left[-\frac{qV_{oc}}{m\,k_B T}\right]\left(\frac{1}{1+\frac{m}{m_\Psi}r_\Psi(V)}\right) \quad (21.a)$$

$$\Psi_{J,dc} = -\Psi_{oc}\exp\left[-\frac{qV_{oc}}{m_\Psi k_B T}\right] \quad (21.b)$$

$$\Psi_{V,dc} = \frac{m_\Psi}{(m_\Psi + m)}\frac{m\,k_B T}{q\,P_{in}} \quad (21.c)$$

$$R_{\Psi,dc} = R_{th}\exp\left[-\frac{qV_{oc}}{mk_B T}\right]\left(\frac{1}{1+\frac{m}{m_\Psi}}\right) \quad (21.d)$$

where the photocurrent resistance factor $r_\Psi$ comes after (S6), resulting $r_\Psi > 1$ for low illumination intensities before $r_\Psi \to 0$ when $V_{oc} \to V_{bi}$.

Note that (21) explains the three main experimental observations. First, the decrease trends of $\Psi_J$ and $\Psi_V$ as $P_{in}$ is augmented at OC in (21b,c) agree with the low frequency limits of IMPS and IMVS spectra, respectively. Second, from the parentheses in (21a,d) we see that $R_{\Psi,dc} \geq R_{dc}$, as the experimental evidence discussed in the previous section. And third, by comparing IS and LIMIS *dc* resistances we realize that they converge only when $m_\Psi \gg m$, so both parentheses in (21a,d) equal unity, and as $V_{oc}$ increases, as suggested more evidently by the SiSC behavior in **Figure 6** and **Figure 11**.

## 3. Conclusions

In summary, the concept and initial theoretical considerations for a new method of characterization of all-solid-state solar cells were presented: the light intensity modulated impedance spectroscopy (LIMIS). Differently to the standard potentiostatic impedance spectroscopic (IS), LIMIS perturbates photo-sensitives samples with light and the photocurrent and photovoltage signals are recorded and analyzed.

Preliminary LIMIS spectra measurements were presented and compared with IS spectra, resulting similar in shape but in most of the cases the total impedance from LIMIS exceeds that from IS. That difference is first analyzed as potential figure of merit for evaluation performance and degradation of solar cells. Those results and the light dependency of the current responsivity at open circuit justified a correction to the empirical Shockley equation, including a bias dependent photo-current term.

Moreover, it has been shown how the total differential resistances and capacitances are reduced and augmented, respectively with respect to IS, illustrating the photoconductivity increase under illumination for the solar cells. This effect corrects the evaluation of the lifetimes, which is a factor to consider in the typical differences when evaluating that parameter by different techniques, like TPV.

# Conflicts of interest

There are no conflicts to declare.

# Acknowledgements

We thank Ministerio de Ciencia, Innovación y Universidades of Spain under project (MAT2016-76892-C3-1-R). O.A. acknowledges the financial support from the VDI/VD Innovation + Technik GmbH (Project-title: PV-ZUM) and the SAOT funded by the German Research Foundation (DFG) in the framework of the German excellence initiative.

# AUTHOR INFORMATION

## Corresponding Author

*E-mail: osbel.almora@fau.de

## ORCID

Osbel Almora: 0000-0002-2523-0203

Germà Garcia-Belmonte: 0000-0002-0172-6175

Christoph J. Brabec: 0000-0002-9440-0253

# *Supporting Information:*

# Light Intensity Modulated Impedance Spectroscopy (LIMIS) in All-Solid-State Solar Cells at Open Circuit

Osbel Almora[1,2,3]*, Yicheng Zhao[1], Xiaoyan Du[1], Thomas Heumueller[1], Gebhard J. Matt[1], Germà Garcia-Belmonte[3] and Christoph J. Brabec[1]

[1]*Institute of Materials for Electronics and Energy Technology (i-MEET), Friedrich-Alexander-Universität Erlangen-Nürnberg, 91058 Erlangen, Germany;*

[2]*Erlangen Graduate School in Advanced Optical Technologies (SAOT), Friedrich-Alexander-Universität Erlangen-Nürnberg, 91052 Erlangen, Germany*

[3]*Institute of Advanced Materials (INAM), Universitat Jaume I, 12006 Castelló, Spain*

*osbel.almora@fau.de

**Table S1:** List of acronyms, symbols and abbreviations

| | | | |
|---|---|---|---|
| 2D | Two-dimensional | $E_g$ | Band-gap energy (eV) |
| 3D | Three-dimensional | $E_i$ | Intrinsic energy level (eV) |
| *ac* | Alternating current (mode) | $E_V$ | Valence band maximum level (eV) |
| $\beta$ | Radiative recombination coefficient (cm$^3\cdot$s$^{-1}$) | ESL | Electron selective layer |
| | | ETL | Electron transport layer |
| $b$ | Power law for the relation *dc* resistance vs. incident light intensity | ETM | Electron transport material |
| | | EQE | External quantum efficiency |
| | | $f$ | Frequency (Hz) |
| $c$ | Speed of light in vacuum (299 792 458 m$\cdot$s$^{-1}$) | $f_\tau$ | Characteristic frequency (Hz) |
| | | FA | Formamidinium |
| $C$ | Capacitance (F$\cdot$cm$^{-2}$) | FF | Fill factor |
| $C^*$ | Complex capacitance (F$\cdot$cm$^{-2}$) | $\tilde{\gamma}, \tilde{\gamma}_1$ | Complex *ac* surface recombination factors |
| $C_{bulk}$ | Bulk capacitance (F$\cdot$cm$^{-2}$) | | |
| CB | Conduction band | $G$ | Generation rate (cm$^{-3}\cdot$s$^{-1}$) |
| $C_{diff}$ | Diffusion layer cap. (F$\cdot$cm$^{-2}$) | $G_0$ | Generation rate at $x=0$ (cm$^{-3}\cdot$s$^{-1}$) |
| $C_{dl}$ | Depletion layer cap. (F$\cdot$cm$^{-2}$) | | |
| $C_g$ | Geometric capacitance (F$\cdot$cm$^{-2}$) | $\tilde{G}$ | Real *ac* perturbation generation rate amplitude (cm$^{-3}\cdot$s$^{-1}$) |
| $C_{Hf}, C_{Lf}$ | High and low frequencies capacitances, respectively, from IS and LIMIS spectra (F$\cdot$cm$^{-2}$) | | |
| | | $\bar{G}$ | Real *dc* generation rate (cm$^{-3}\cdot$s$^{-1}$) |
| | | $h$ | Planck's constant (6.626×10$^{-34}$ J$\cdot$s) |
| $C_{Hl}$ | Helmholtz layer capacitance (F$\cdot$cm$^{-2}$) | HSL | Hole selective layer |
| | | HTL | Hole transport layer |
| $C_{IS}$ | Capacitance from IS (F$\cdot$cm$^{-2}$) | HTM | Hole transport material |
| $C_\Psi$ | Capacitance from LIMIS (F$\cdot$cm$^{-2}$) | $i$ | Imaginary number ($\sqrt{-1}$) |
| $C_\mu$ | Chemical capacitance (F$\cdot$cm$^{-2}$) | IMPS | Intensity modulated photocurrent spectroscopy |
| *dc* | Direct current (mode) | IMVS | Intensity modulated photovoltage spectroscopy |
| DD | Drift-diffusion | | |
| DFT | Density function theory | IHys | Inverted hysteresis |
| $\mathfrak{D}$ | Electric displacement (C$\cdot$cm$^{-2}$) | IS | Impedance spectroscopy |
| $\tilde{\delta}$ | Complex *ac* difference factor between IS and LIMIS | $J$ | Current density (A$\cdot$cm$^{-2}$) |
| | | $\tilde{J}$ | Complex *ac* current density signal amplitude (A$\cdot$cm$^{-2}$) |
| $D$ | Diffusion coefficient (cm$^2\cdot$s$^{-1}$) | | |
| $D_n, D_p$ | Diffusion coefficient for electrons and holes, respectively (cm$^2\cdot$s$^{-1}$) | $\bar{J}$ | Real *dc* current density (A$\cdot$cm$^{-2}$) |
| | | $J_n$ | Electron current density (A$\cdot$cm$^{-2}$) |
| | | $J_p$ | Holes current density (A$\cdot$cm$^{-2}$) |
| $\Delta Z_\Psi$ | Complex normalized photo-impedance difference LIMIS-IS | $J_{ph}$ | Photocurrent density (A$\cdot$cm$^{-2}$) |
| $\Delta Z_\Psi'$ | normalized difference of real parts of photo-impedance LIMIS-IS | $J_s$ | Reverse bias diode dark saturation current density (A$\cdot$cm$^{-2}$) |
| | | $J_{sc}$ | Short-circuit current density (A$\cdot$cm$^{-2}$) |
| $\varepsilon$ | Dielectric constant | | |
| $\varepsilon_0$ | Vacuum permittivity (8.85×10$^{14}$ F$\cdot$cm$^{-1}$) | $J-V$ | Current density-voltage characteristic (plane) |
| $E$ | Energy (eV or J) | $k_B$ | Boltzmann constant (1.38×10$^{-23}$ J$\cdot$K$^{-1}$) |
| EC | Equivalent circuit | | |
| $E_C$ | Conduction band minimum energy level (eV) | $\lambda$ | Photon wavelength (nm) |
| | | $L$ | Distance between electrodes/ Distance between selective contacts larger than $w+L_D$ (cm) |
| $E_{Fn}, E_{Fp}$ | Quasi-Fermi level of electrons and holes, respectively (eV) | | |

| Symbol | Description | Symbol | Description |
|---|---|---|---|
| $L_{bulk}$ | Thickness of the absorber bulk layer (cm) | $N_V$ | Effective density of states at the valence band (cm$^{-3}$) |
| $L_d$ | Diffusion length (cm) | $\omega$ | Angular frequency (rad·s$^{-1}$) |
| $\tilde{L}_d$ | Complex *ac* diffusion length signal amplitude (cm) | $\omega_0$ | Characteristic recombination frequency (rad·s$^{-1}$) |
| $\bar{L}_d$ | Real *dc* diffusion length (cm) | $\omega_\beta$ | Characteristic radiative recombination frequency (rad·s$^{-1}$) |
| $L_D$ | Debye length (cm) | OC | Open-circuit |
| LIMIS | Light intensity modulated impedance spectroscopy | OrgSCs | Organic solar cells |
| LIMTAS | Light intensity modulated thermal admittance spectroscopy | $\varphi$ | Electrostatic Potential (V) |
| $\mu$ | Electronic mobility (cm$^2$·V$^{-1}$·s$^{-1}$) | $\phi$ | Phase shift from IS (rad) |
| $\mu_n, \mu_p$ | Electrons and holes mobilities, respectively (cm$^2$·V$^{-1}$·s$^{-1}$) | $\phi_J$ | Phase shift from IMPS (rad) |
| $m$ | Diode ideality factor | $\phi_n$ | Phase shift of the *ac* minority carriers signal amplitude (rad) |
| $m_C$ | Capacitance ideality factor | $\phi_V$ | Phase shift from IMVS (rad) |
| $m_\Psi$ | Photocurrent ideality factor | $\phi_\Psi$ | Phase shift from LIMIS (rad) |
| MA | Methylammonium | $p$ | Holes charge density (cm$^{-3}$) |
| MAPI | CH$_3$NH$_3$PbI$_3$ | $p_0$ | Real *dc* dark equilibrium minority carrier holes charge density (cm$^{-3}$) |
| $n$ | Electron charge density/ Average minority carriers charge density (cm$^{-3}$) | PCE | Power conversion efficiency |
| | | PC | Photocurrent (A·cm$^{-2}$) |
| | | $P_{in}$ | Light incident power (W·cm$^{-2}$) |
| $\tilde{n}$ | Complex *ac* average minority carriers charge density signal amplitude (cm$^{-3}$) | $\tilde{P}_{in}$ | Real *ac* light incident power perturbation amplitude (W·cm$^{-2}$) |
| $\bar{n}$ | Real *dc* steady-state over-equilibrium minority carriers charge density (cm$^{-3}$) | $\bar{P}_{in}$ | Real *dc* light incident power density (W·cm$^{-2}$) |
| | | $\Psi_J$ | Current responsivity/ Complex current responsivity transfer function (A·W$^{-1}$) |
| $n_0$ | Real *dc* dark equilibrium minority carriers charge density (cm$^{-3}$) | $\Psi_{J,dc}$ | Photo-current responsivity from the *dc* $J-V$ curve (A·W$^{-1}$) |
| $\bar{n}_0$ | Total real *dc* average minority carriers charge density (cm$^{-3}$) | $\Psi_J', \Psi_J''$ | Real and imaginary parts of $\Psi_J$ (A·W$^{-1}$) |
| $N_\mu$ | Effective total equilibrium charge density that contributes to chemical capacitance (cm$^{-3}$) | $\Psi_{sc}$ | Real bias-independent current responsivity at SC (A·W$^{-1}$) |
| $N_A$ | Ionized fixed acceptor doping concentration (cm$^{-3}$) | $\Psi_{oc}$ | Real bias-independent current responsivity at OC (A·W$^{-1}$) |
| $N_C$ | Effective density of states at the conduction band (cm$^{-3}$) | $\Psi_V$ | Voltage responsivity/ Complex voltage responsivity transfer function (V·W$^{-1}$·cm$^2$) |
| $N_{CV}$ | Average effective density of states at CB and VB (cm$^{-3}$) | | |
| $N_D$ | Ionized fixed donor doping concentration (cm$^{-3}$) | $\Psi_{V,dc}$ | Photo-voltage responsivity from the *dc* $J-V$ curve (V·W$^{-1}$·cm$^2$) |
| $N_{eff}$ | Effective concentration of fixed ionized species in the depletion zone: $N_D$ or $N_A$ (cm$^{-3}$) | $\Psi_V', \Psi_V''$ | Real and imaginary parts of $\Psi_V$ (V·W$^{-1}$·cm$^2$) |
| $N_{ion}$ | Average concentration of ionized charge (cm$^{-3}$) | PSCs | Perovskite solar cells |
| | | PV | Photovoltaic |
| $N_\mu$ | Effective total equilibrium charge density that contributes to chemical capacitance (cm$^{-3}$) | $q$ | Elementary charge (1.6×10$^{-19}$ C) |
| | | $Q$ | Charge density (C·cm$^{-2}$) |
| | | $\rho$ | Charge density (C·cm$^{-3}$) |
| $R$ | Resistance (Ω·cm²) | $TAS$ | Thermal admittance spectroscopy |
| $R_{bulk}$ | Bulk resistance (Ω·cm²) | TPV | Transient photovoltage |
| $R_{dc}$ | $dc$ resistance from $J-V$ curve partial derivative (Ω·cm²) | $U$ | Recombination rate (cm⁻³·s⁻¹) |
| | | $\hat{v}_{1,2,3}$ | Unitary direction vectors |
| $R_{IS}$ | Resistance from IS (Ω·cm²) | $V$ | Voltage (V) |
| $r_\Psi$ | Photocurrent resistance factor | $\widetilde{V}$ | Real $ac$ voltage perturbation amplitude (V) |
| $R_\Psi$ | Resistance from LIMIS (Ω·cm²) | | |
| $R_{\Psi,dc}$ | LIMIS resistance from $dc\ J-V$ curves (Ω·cm²) | $\overline{V}$ | Real $dc$ voltage (V) |
| | | $V_{bi}$ | Built-in voltage (V) |
| $R_T$ | Total $C$-coupled resistance (Ω·cm²) | VB | Valence band |
| | | $V_\varsigma$ | collection threshold voltage (V) |
| $R_{th}$ | Thermal recombination resistance (Ω·cm²) | $V_\Psi$ | Photocurrent resistance voltage |
| | | $V_{oc}$ | Open circuit voltage (V) |
| $R_{series}$ | Series resistance (Ω·cm²) | $\widetilde{V}_{oc}$ | Complex $ac$ open circuit voltage signal amplitude (V) |
| $R_{sh}$ | Shunt resistance (Ω·cm²) | | |
| $\varsigma$ | Collection fraction | $\overline{V}_{oc}$ | Real $dc$ open circuit voltage (V) |
| SC | Short-circuit | $w$ | Depletion layer width (cm) |
| SiSCs | Silicon solar cells | $\widetilde{w}$ | Complex $ac$ depletion layer width modulated amplitude (cm) |
| $S_r$ | Surface recombination velocity (cm·s⁻¹) | | |
| | | $\overline{w}$ | Real $dc$ depletion layer width (cm) |
| $S_{rn}, S_{rp}$ | Surface recombination velocity of electrons and holes, respectively (cm·s⁻¹) | $\xi$ | Electric field (V·cm⁻¹) |
| | | $x$ | Distance from the interface (cm) |
| | | $Z$ | Impedance/ Complex impedance transfer function from IS (Ω·cm²) |
| $\tau$ | Lifetime/ Lifetime from TPV/ Non-radiative recombination lifetime/ Characteristic RC time constant from IS and LIMIS (s) | | |
| | | $Z', Z''$ | Real and imaginary parts of $Z$, respectively (Ω·cm²) |
| $\tau_{Hf}, \tau_{Lf}$ | High and low frequencies characteristic RC time constants from IS and LIMIS (s) | $Z_T'$ | Total or low frequency limit of real part of impedance (Ω·cm²) |
| | | $Z_\Psi$ | Photo-impedance from LIMIS/ Complex photo-impedance transfer function from LIMIS (Ω·cm²) |
| $\mathfrak{I}_{-1/2}$ | Fermi-Dirac 1/2 integral | | |
| $t$ | Time (s) | | |
| $T$ | Temperature (K) | | |

# S1. Experimental

## S1.1. Devices structures and performance

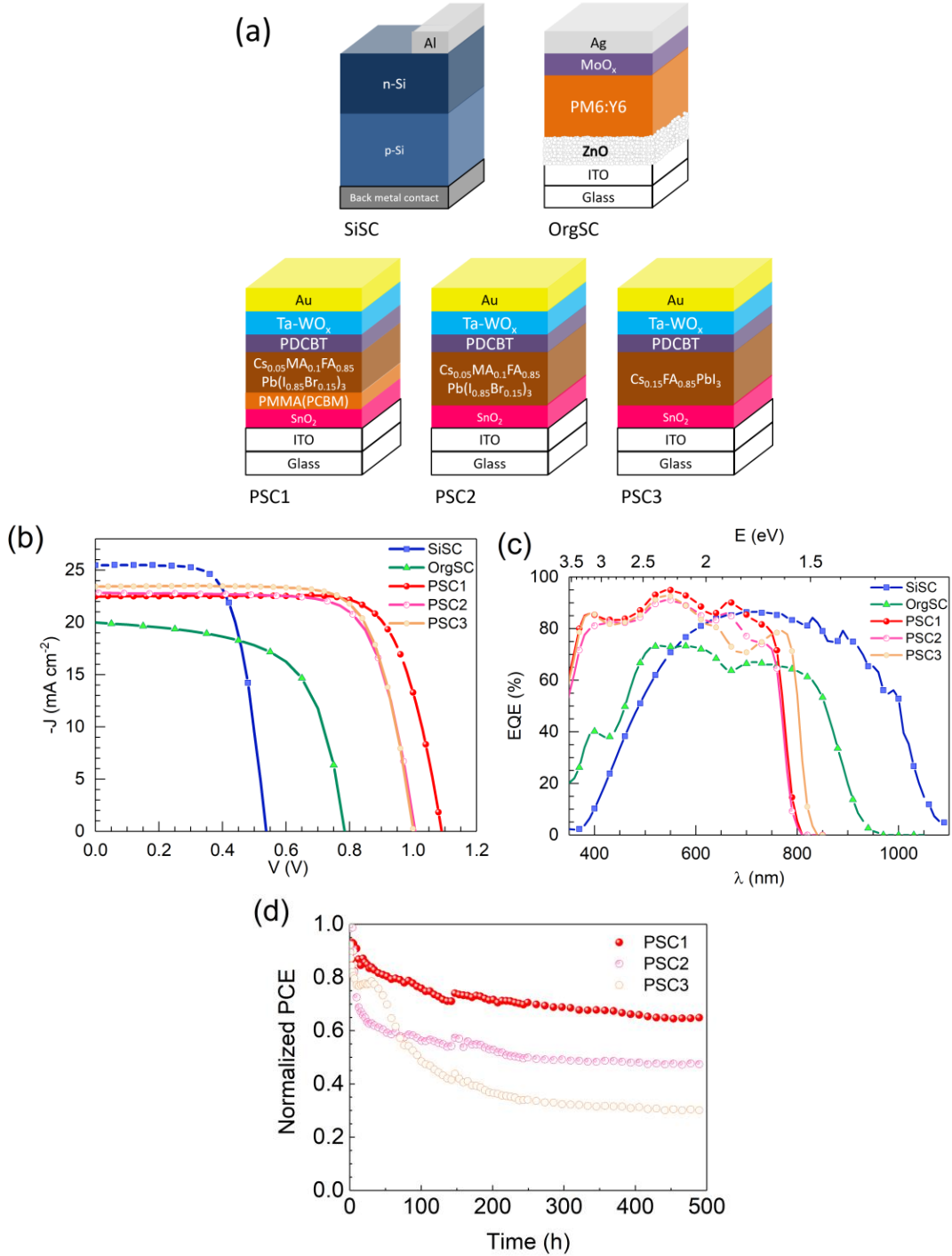

**Figure S1.** (a) Schemed structures with labels and experimental (b) current voltages characteristics under 1 sun standard AM1.5G illumination, (c) external quantum efficiency (EQE) for the studied devices. Performance parameters are summarized in **Table S2**. (d) Illustrative time evolution of normalized power conversion efficiency

(PCE) of the perovskite devices showing most of degradation occurring in the first 200 h under 1 sun white LED illumination in $N_2$ atmosphere.

**Table S2:** Performance parameters from the measured devices measured under 100 mW cm$^{-2}$ light intensity from a Newport AAA AM1.5G solar simulator. Here $J_{sc}$ is the short-circuit current density, $V_{oc}$ the open-circuit voltage, $FF$ de fill factor and $PCE$ the power conversion efficiency.

| Label | Structure | Area cm$^2$ | $J_{sc}$ mA·cm$^{-2}$ | $V_{oc}$ mV | $FF$ % | $PCE$ % |
|---|---|---|---|---|---|---|
| SiSC | n-Si/p-Si | 4.0 | 25.5 | 538 | 67.7 | 9.3 |
| OrgSC | ITO/ZnO/PM6:Y6/MoOx/Ag | 0.1 | 20.0 | 787 | 62.0 | 9.7 |
| PSC1 | ITO/SnO$_2$/ PMMA(PCBM)/ Cs$_{0.05}$MA$_{0.1}$FA$_{0.85}$Pb(I$_{0.85}$Br$_{0.15}$)$_3$/ PDCBT/Ta-WO$_x$/Au | 0.1 | 22.5 | 1090 | 74.8 | 18.3 |
| PSC2 | ITO/SnO$_2$/Cs$_{0.05}$MA$_{0.1}$FA$_{0.85}$Pb(I$_{0.85}$Br$_{0.15}$)$_3$/ PDCBT/ Ta-WO$_x$ /Au | 0.1 | 23.3 | 1007 | 71.9 | 16.8 |
| PSC3 | ITO/SnO$_2$/Cs$_{0.15}$FA$_{0.85}$PbI$_3$/PDCBT/ Ta-WO$_x$ /Au | 0.1 | 23.4 | 1000 | 74.4 | 17.4 |

## S1.2. Silicon solar cell

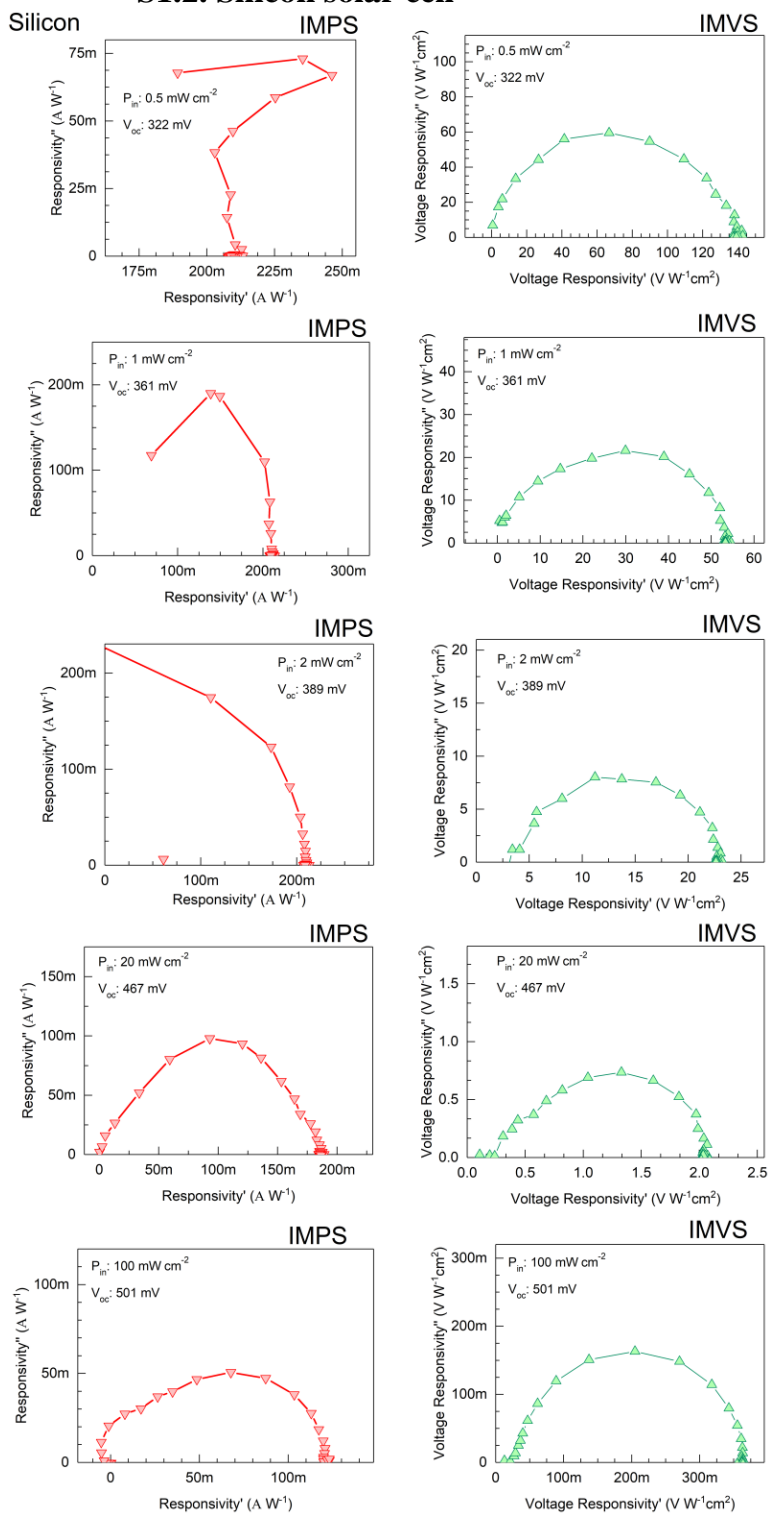

**Figure S2.** Absolute IMPS and IMVS spectra from the Zahner reference **silicon solar cell** (SiSC) at open circuit under different illumination intensities, as indicated.

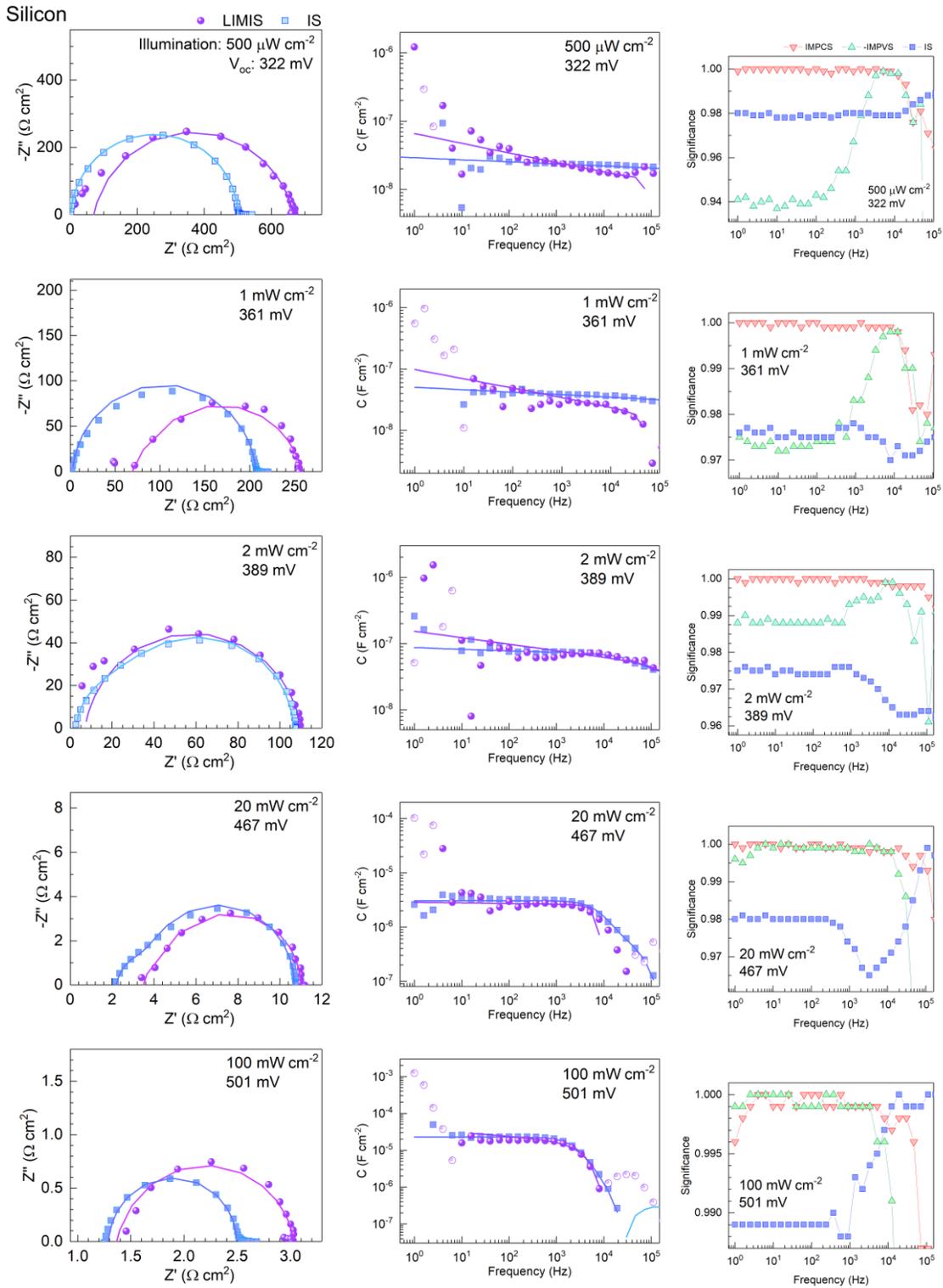

**Figure S3.** LIMIS and IS spectra from the Zahner reference **silicon solar cell** (SiSC) at open circuit under different illumination intensities, as indicated. Left, central and right panels show impedance Nyquist plots, capacitance Bode plots and significance Bode plots, respectively. In left and central panels, the dots represent the experimental data and the lines are the numerical simulation using the equivalent circuit model of Figure 4b in the main manuscript (only the two higher illumination intensities for IS needed the model of Figure 4b). The empty dots represent negative values.

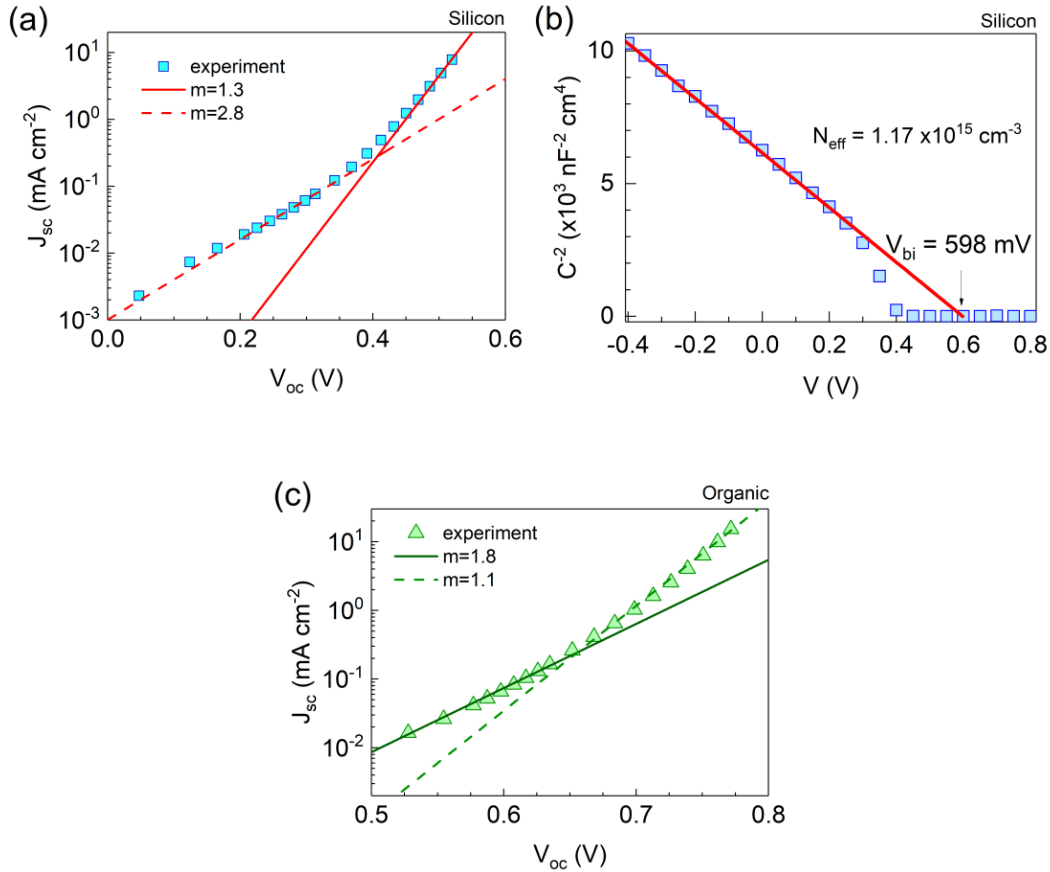

**Figure S4.** (a) Photocurrent-Photovoltage curve and (b) dark Mott-Schottky plot at 10 kHz from the Zahner reference **silicon solar cell** (SiSC). (c) Photocurrent-Photovoltage curve for the **organic solar cell** (OrgSC). Dots are the experimental data and lines are the fittings. Illumination in (a) and (c) was made with a white LED source.

### S1.3. Organic solar cell

*Device fabrication:* The solar cells were fabricated in an inverted architecture (ITO/ZnO/active layer/MoOx/Ag) on ITO-coated glasses using spin-coating in a nitrogen atmosphere. Pre-structured ITO-coated glass was cleaned in sequence with water, acetone and IPA for 10 min. After drying using a nitrogen gun, the substrate was coated with ~30 nm ZnO and dried at 80 ℃ for 5 min. The active layer was spin coated atop ZnO. For all the active layers, chloroform-based solution (16 mg mL$^{-1}$ in total) with donor to acceptor weight ratio of 1:1.2 was used. The active layer thickness is around 100 nm. The active layers were annealed at 110 ℃ for 10 min. 15 nm MoOx and 100 nm Ag was thermally evaporated with a shadow mask subsequently.

Materials: PM6 (batch No. SX8045B100) and Y6 (batch No. DW4132P) were received from 1-Materials. ZnO (Product N-10) was received from Nanograde.

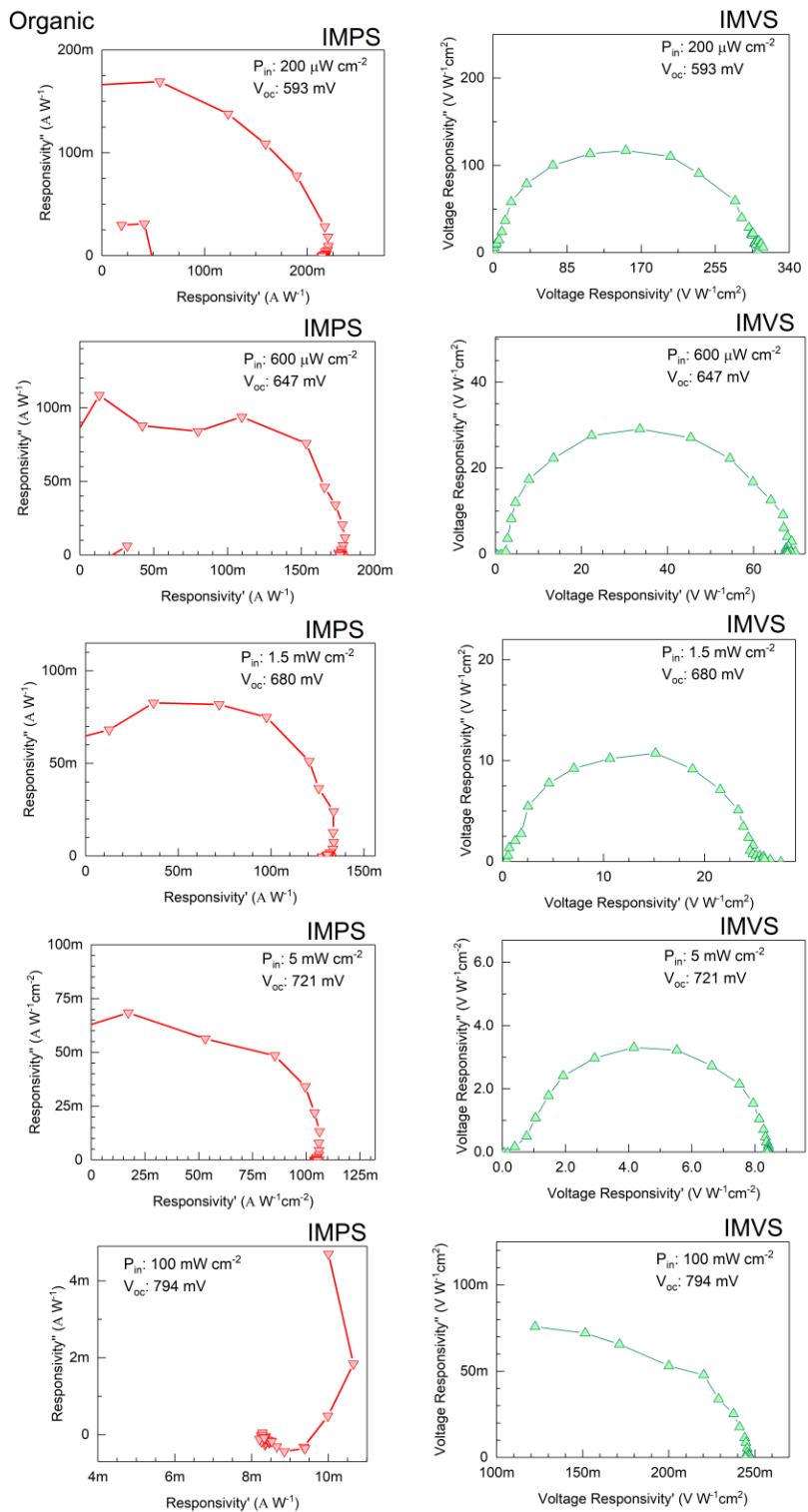

**Figure S5.** Absolute IMPS and IMVS spectra from the **organic solar cell** (OrgSC) at open circuit under different illumination intensities, as indicated.

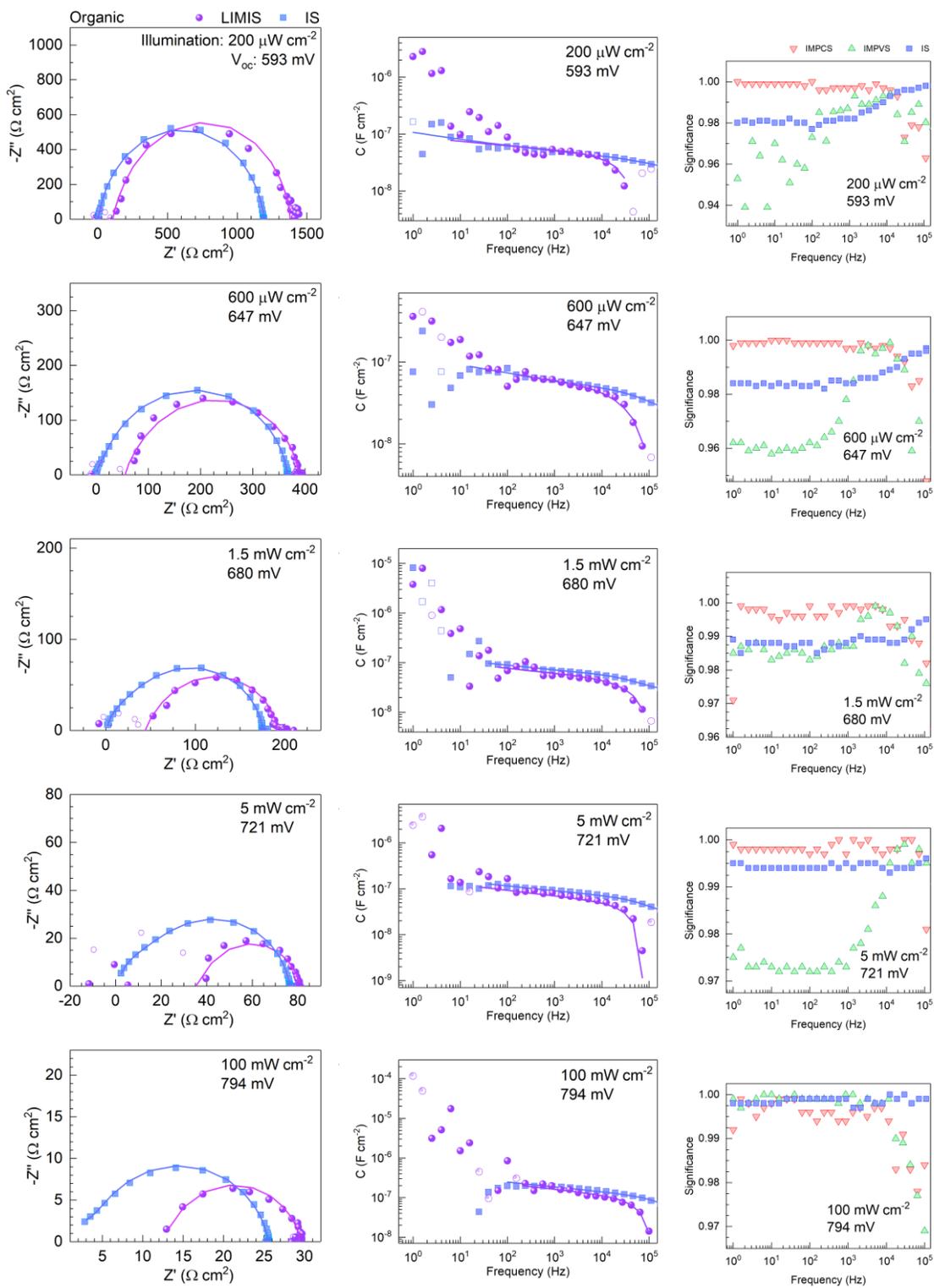

**Figure S6.** LIMIS and IS spectra from the **organic solar cell** (OrgSC) at open circuit under different illumination intensities, as indicated. Left, central and right panels show impedance Nyquist plots, capacitance Bode plots and significance Bode plots, respectively. In left and central panels, the dots represent the experimental data and the lines are the numerical simulation using the equivalent circuit model of Figure 4b. The empty dots represent negative values.

### S1.4. Perovskite solar cell 1

$SnO_2$/PMMA(PCBM)/$Cs_{0.05}MA_{0.1}FA_{0.85}Pb(I_{0.85}Br_{0.15})_3$/PDCBT/Ta-$WO_x$/Au

*Device Fabrication:* Unless stated otherwise, all materials were purchased from Sigma Aldrich or Merck and used as Received. Pre-patterned indium tin oxide (ITO) coated glass was sequentially cleaned using detergent, acetone, and isopropanol. $SnO_2$ nanocrystal solutions were purchased from Alfa company (15% in $H_2O$ colloidal dispersion). $SnO_2$ diluted solution (0.3 mL with 0.9 mL water and 0.9 mL iso-propanol) was spin-coated on the ITO substrates at 3500 rpm for 30s and were then annealed on a hot plate at 150°C for 10 minutes in ambient air. The substrates were immediately transferred to the $N_2$-filled glovebox after cooling. For the PCBM/PMMA mixed solution, 3 mg PCBM and 1 mg PMMA was dissolved into CB solvent, stirring overnight at 50°C. The PCBM/PMMA solution is then spin-coated onto the $SnO_2$ substrate with 5000 rpm for 30s.

The $Cs_{0.05}FA0_{.81}MA_{0.14}PbI_{2.55}Br_{0.45}$ precursor solution was prepared in DMF/DMSO (4/1 V/V) solvent, which contains $PbI_2$ (2.38 M), $PbBr_2$ (0.42 M), FAI (2.26 M), MABr (0.4 M) and CsI (0.14 M). The perovskite films were deposited using a two-step program at 2000 and 5000 r.p.m for 10 and 40 s respectively. During the second step, 200 µL of chlorobenzene was dropped on the spinning substrate at 20 s before the end of the process. After spin-coating, the films were annealed at 100°C for 20 minutes and 150°C for 10 minutes.

PDCBT were dissolved in chlorobenzene at concentrations of 10 mg/mL. It was spin-coated on perovskite film at 2000 rpm for 30 seconds and annealled at 80 °C for 10 minutes. Ta-$WO_x$ was coated onto polymer PDCBT at a speed of 2000 r.p.m for 30s. Finally, a 120-nm-thick Au electrode was deposited onto a hole-transporting layer through a shadow mask by thermal evaporation.

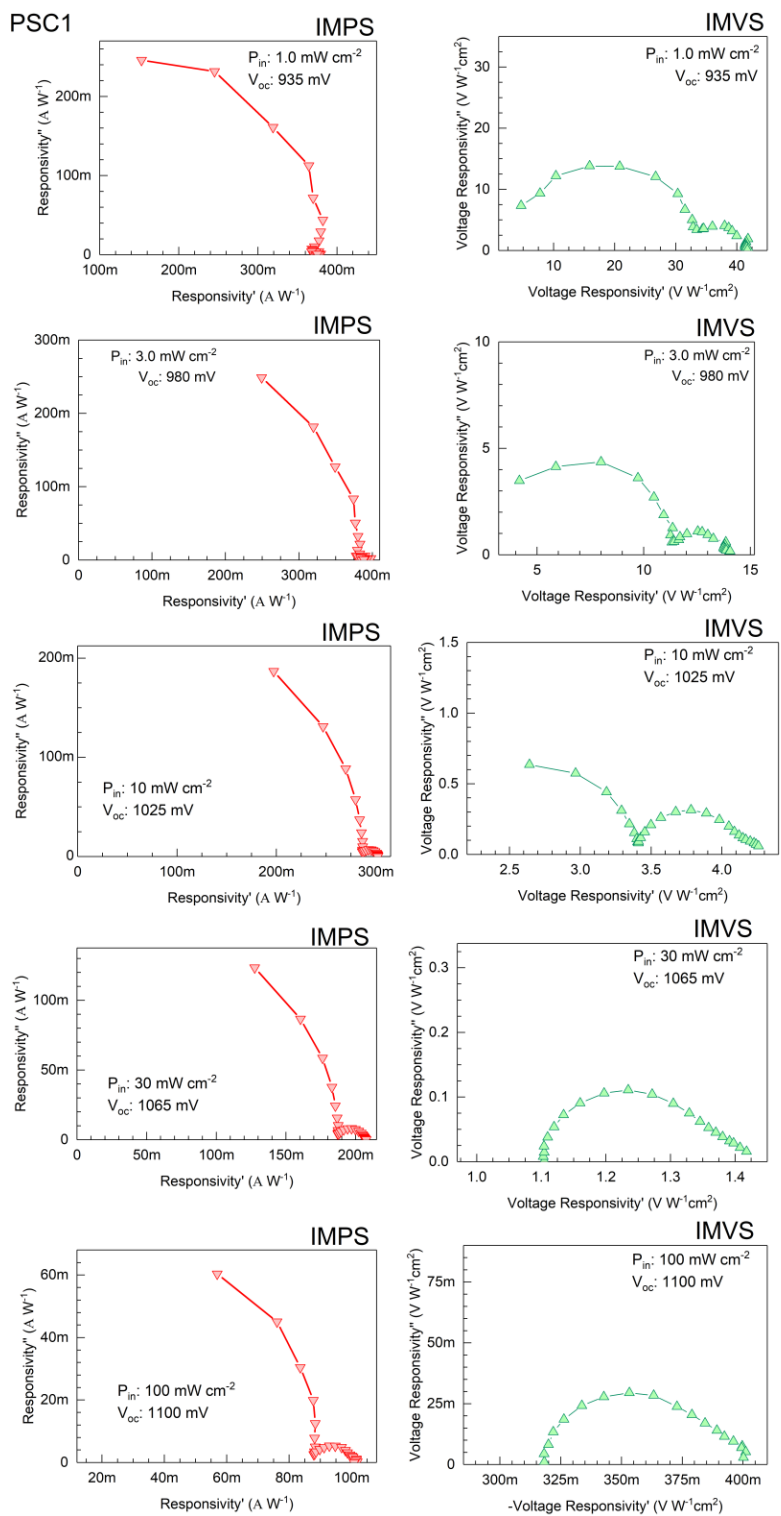

**Figure S7.** Absolute IMPS and IMVS spectra from the SnO$_2$/PMMA(PCBM)/Cs$_{0.15}$FA$_{0.85}$PbI$_3$/PDCBT/WO$_x$/Au **perovskite solar cell** (PSC1) at open circuit under different illumination intensities, as indicated.

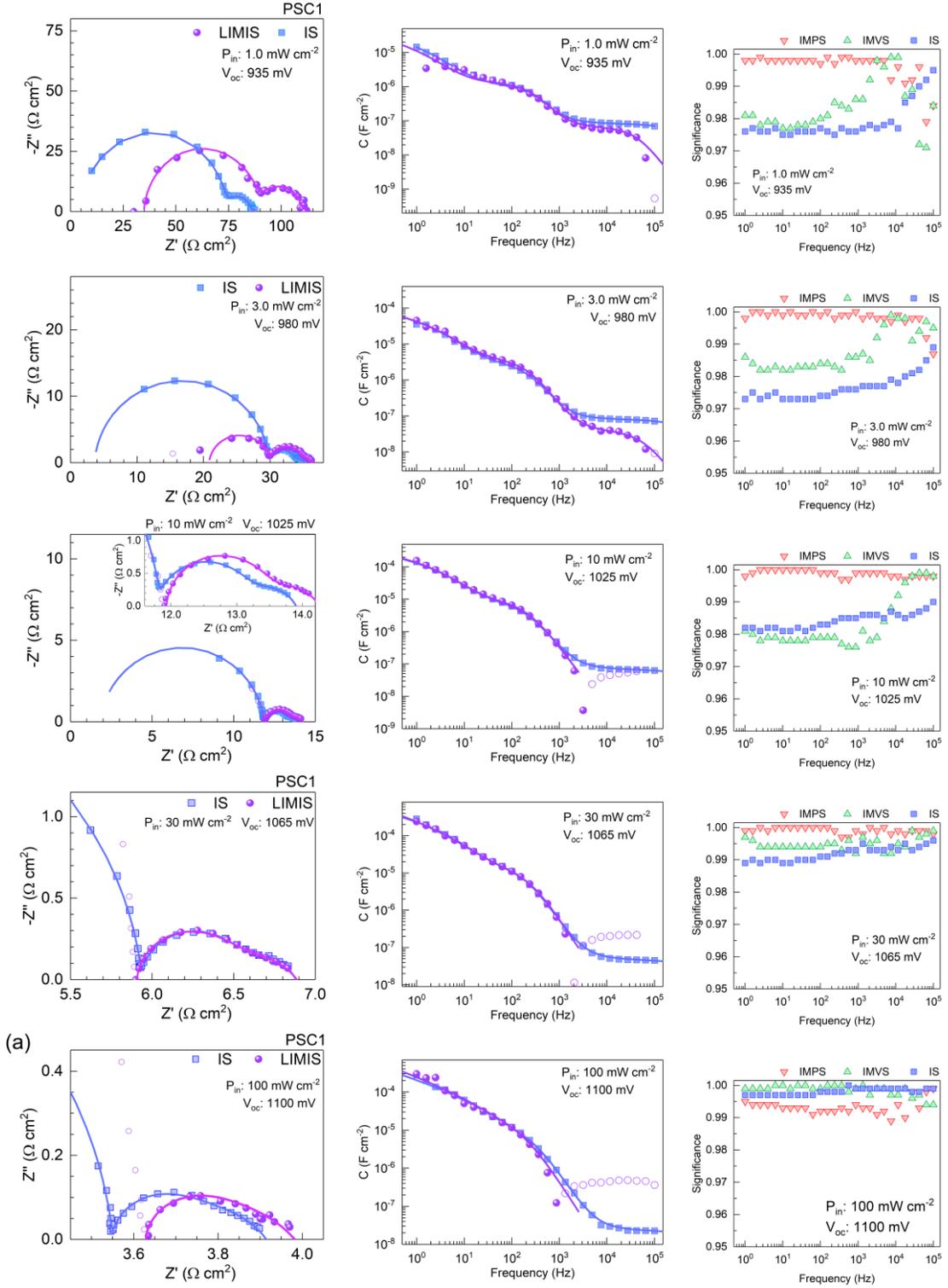

**Figure S8.** LIMIS and IS spectra from the SnO$_2$/PMMA(PCBM)/Cs$_{0.15}$FA$_{0.85}$PbI$_3$/PDCBT/WO$_x$/Au **perovskite solar cell** (PSC1) at open circuit under different illumination intensities, as indicated. Left, central and right panels show impedance Nyquist plots, capacitance Bode plots and significance Bode plots, respectively. In left and central panels, the dots represent the experimental data and the lines are the numerical simulation using the equivalent circuit model of Figure 4c. The empty dots represent negative values.

## S1.5. Perovskite solar cell 2

$SnO_2/Cs_{0.05}MA_{0.1}FA_{0.85}Pb(I_{0.85}Br_{0.15})_3/PDCBT/WO_x/Au$

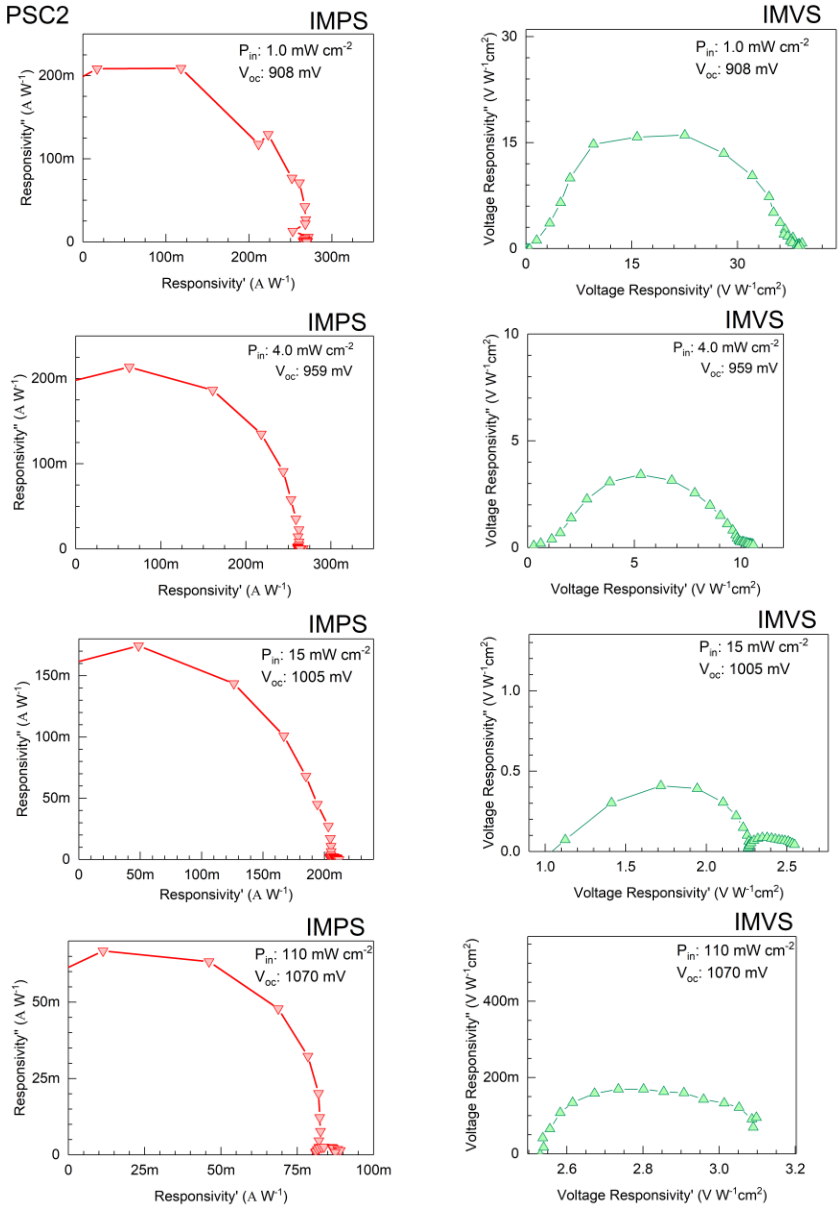

**Figure S9.** Absolute IMPS and IMVS spectra from the Zahner reference $SnO_2/Cs_{0.05}MA_{0.1}FA_{0.85}Pb(I_{0.85}Br_{0.15})_3/PDCBT/WO_x/Au$ **perovskite solar cell** (PSC-2) at open circuit under different illumination intensities, as indicated.

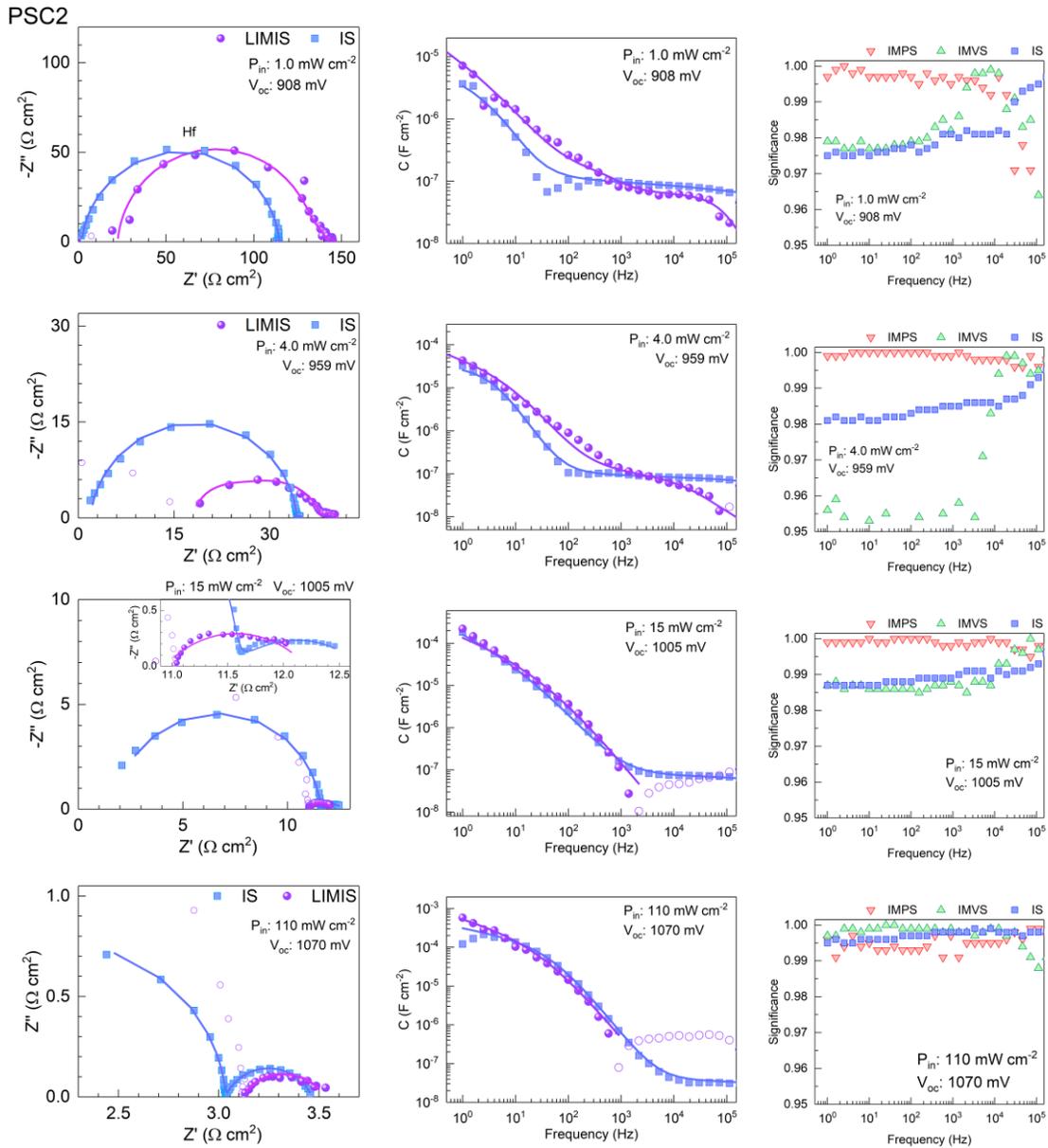

**Figure S10.** LIMIS and IS spectra from the SnO$_2$/Cs$_{0.05}$MA$_{0.1}$FA$_{0.85}$Pb(I$_{0.85}$Br$_{0.15}$)$_3$/PDCBT/WO$_x$/Au **perovskite solar cell** (PSC2) at open circuit under different illumination intensities, as indicated. Left, central and right panels show impedance Nyquist plots, capacitance Bode plots and significance Bode plots, respectively. In left and central panels, the dots represent the experimental data and the lines are the numerical simulation using the equivalent circuit model of Figure 4b. The empty dots represent negative values.

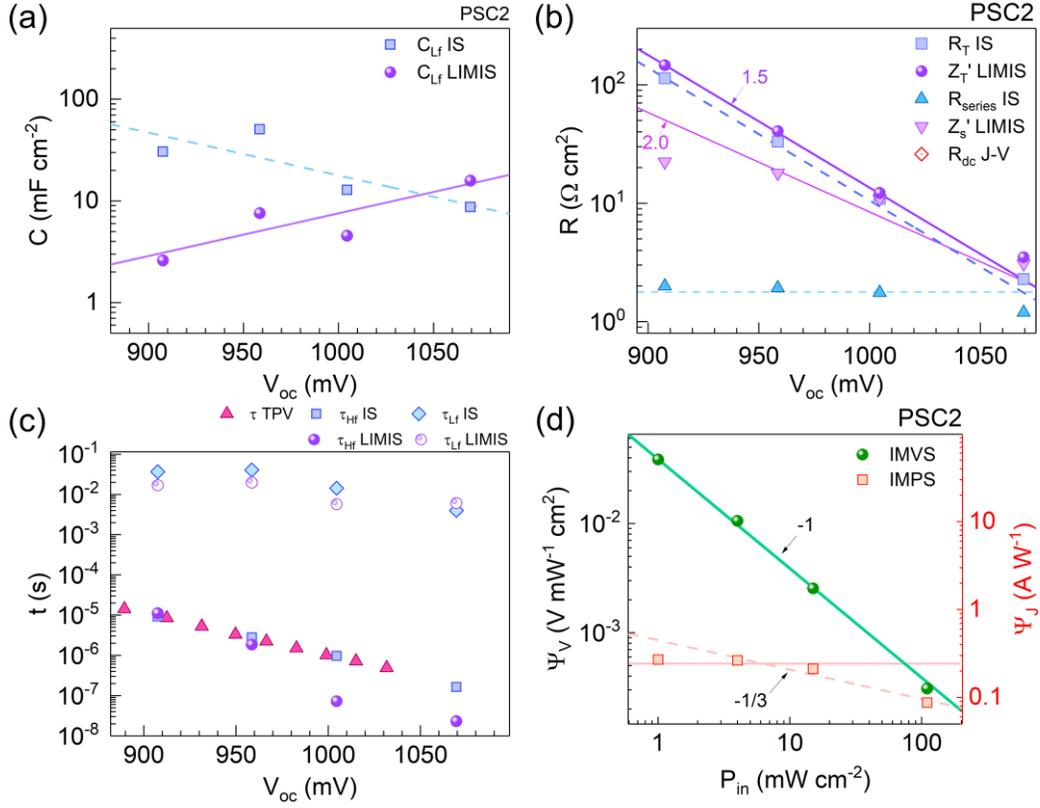

**Figure S11.** Perovskite solar cell (PSC2) numerical simulation (lines in **Figure S10**) results from IS and LIMIS experimental data (dots in **Figure S10**): (a) capacitance, (b) resistance and (c) characteristic times from LIMIS and IS (EC model in **Figure 7**c) and TPV (mono-exponential decay model). The lines are the fittings to the $R_{dc}$ behavior (15.a) with $m$ as indicated with arrows in (b) and in (a) the lines are exponential fittings to $C = C_1 \exp[qV_{oc}/m_C k_B T]$ with $m_C = 4$, and $C_1 = 47.8\ \mu\text{F} \cdot \text{cm}^{-2}$ for LIMIS and $m_C = -4$, and $C_1 = 283\ \text{F} \cdot \text{cm}^{-2}$ for IS. (c) Low frequency limits of the voltage and current responsivities for different light intensities (see **Figure S9**) where the solid lines belong to fitting to $\Psi_V \propto P_{in}^{-1}$ and $\Psi_J \propto P_{in}^{0}$ and the dashed one to $\Psi_J \propto P_{in}^{-1/3}$

## S1.6. Perovskite solar cell 3
$SnO_2/Cs_{0.15}FA_{0.85}PbI_3/PDCBT/WO_x/Au$

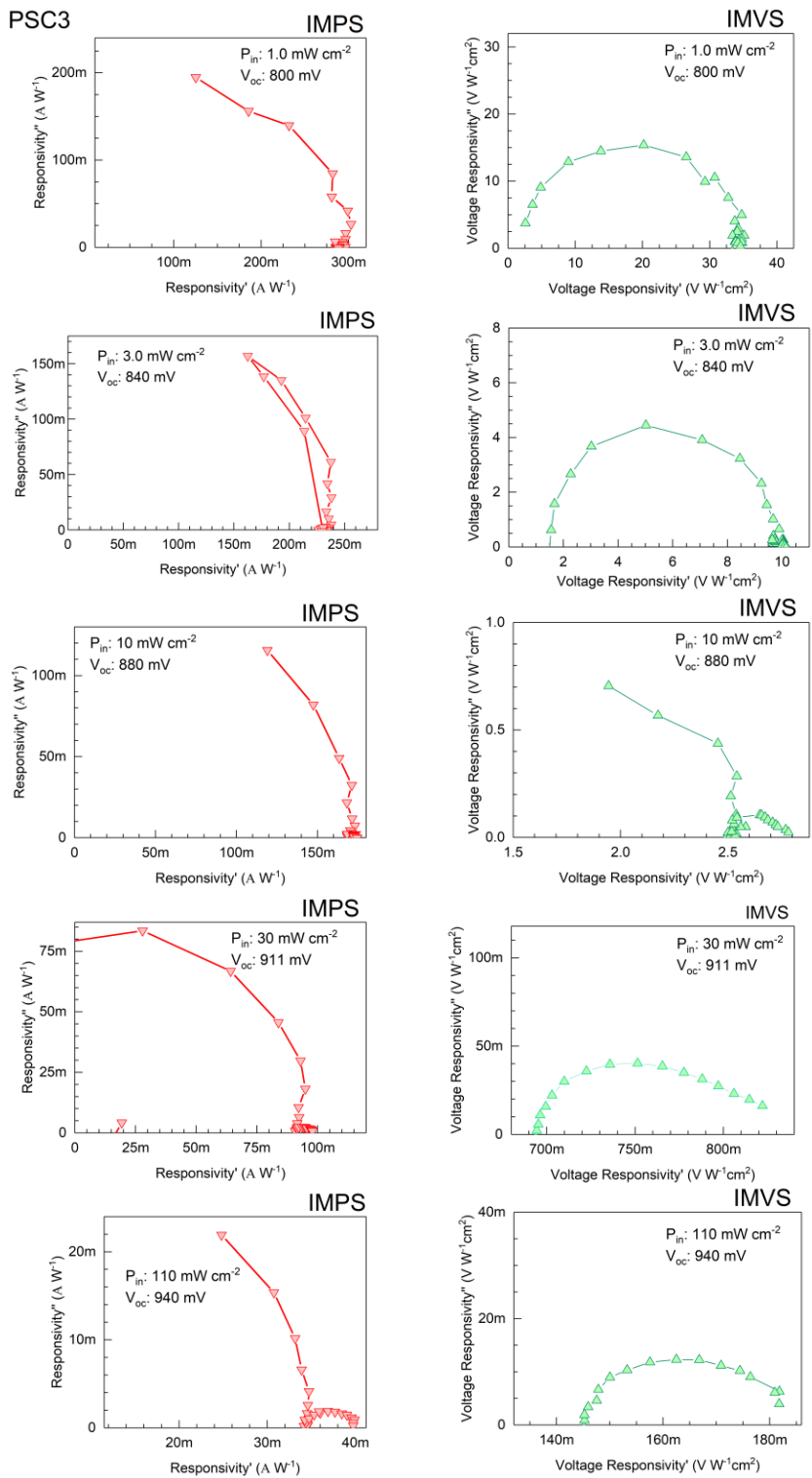

**Figure S12.** Absolute IMPS and IMVS spectra from the $SnO_2/Cs_{0.15}FA_{0.85}PbI_3/PDCBT/WO_x/Au$ **perovskite solar cell** (PSC3) at open circuit under different illumination intensities, as indicated.

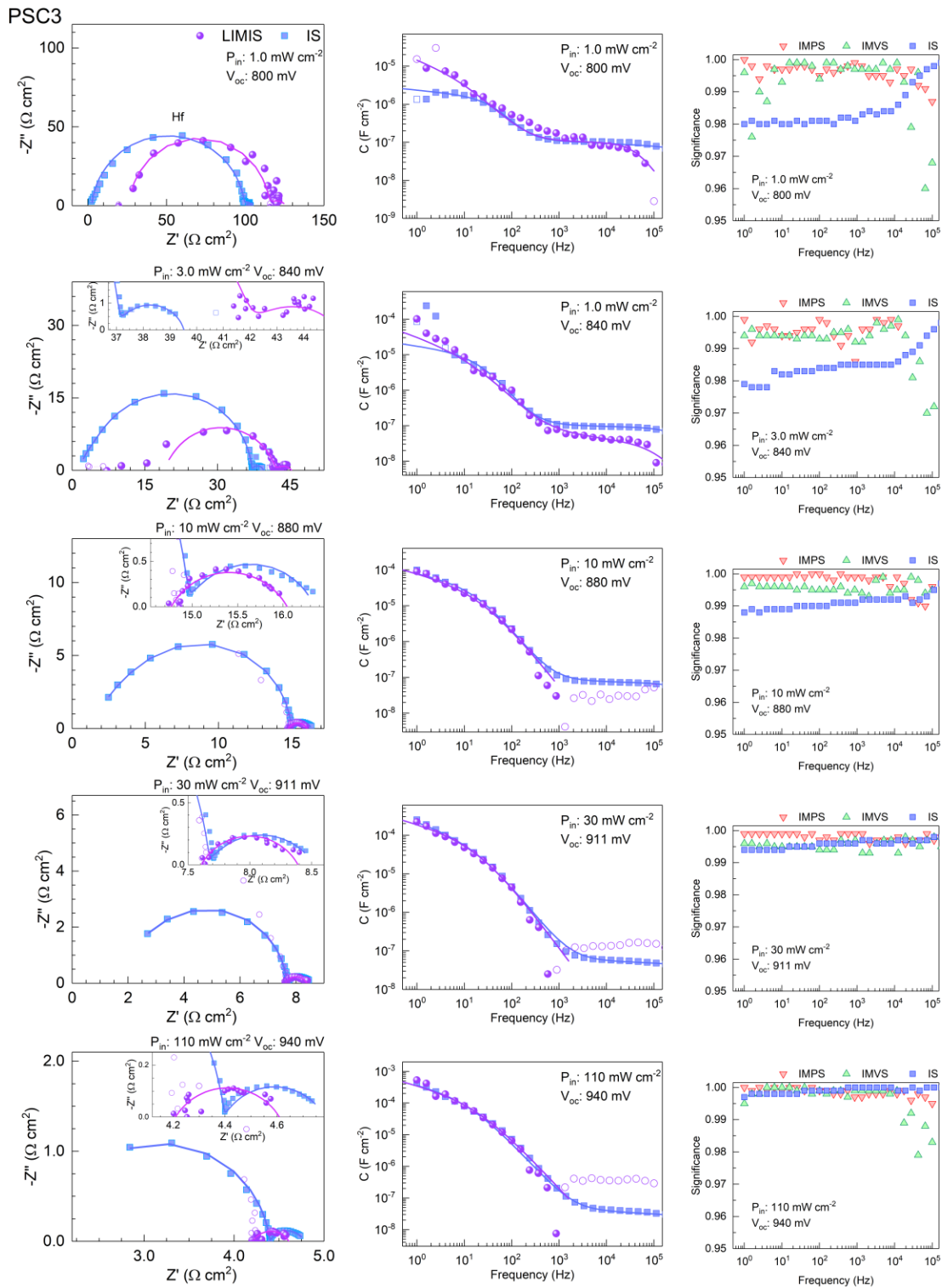

**Figure S13.** LIMIS and IS spectra from the SnO$_2$/Cs$_{0.15}$FA$_{0.85}$PbI$_3$/PDCBT/WO$_x$/Au **perovskite solar cell** (PSC3) at open circuit under different illumination intensities, as indicated. Left, central and right panels show impedance Nyquist plots, capacitance Bode

plots and significance Bode plots, respectively. In left and central panels, the dots represent the experimental data and the lines are the numerical simulation using the equivalent circuit model of Figure 4b. The empty dots represent negative values.

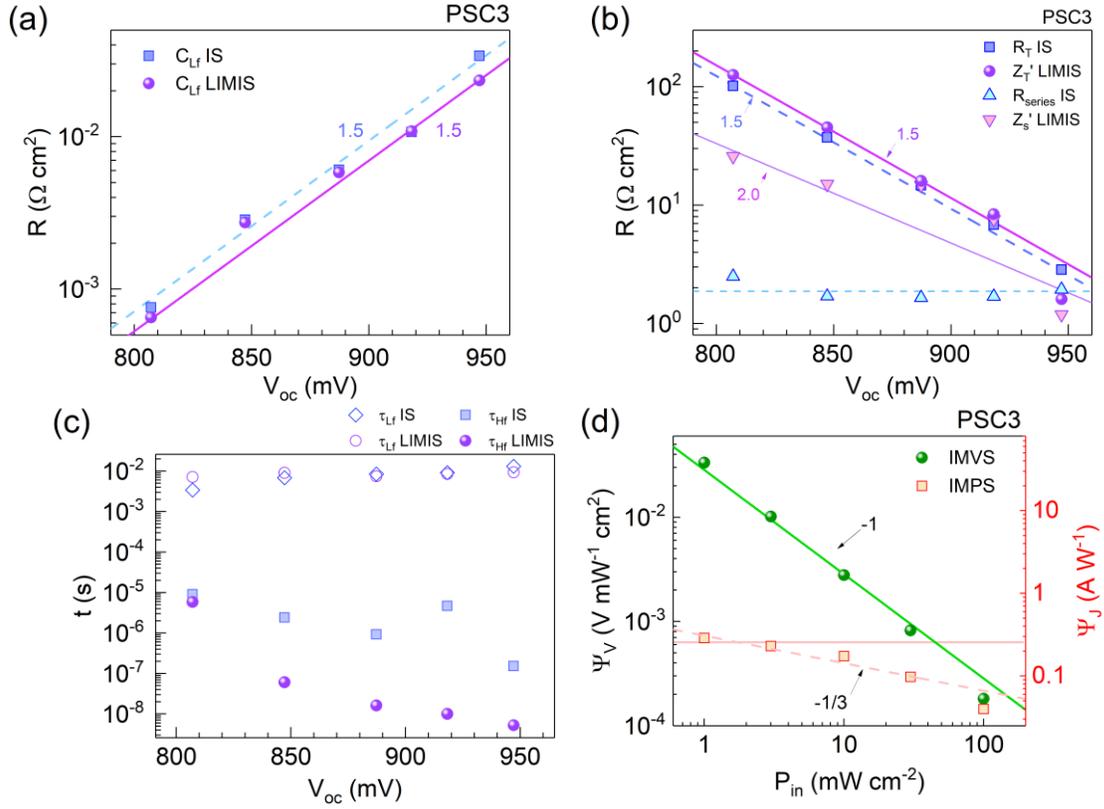

**Figure S14.** Perovskite solar cell (PSC3) numerical simulation results (lines in **Figure S13**) from IS and LIMIS experimental data (dots in **Figure S13**): (a) capacitance, (b) resistance and (c) characteristic times from LIMIS and IS (EC model in Figure 7c) and TPV (mono-exponential decay model). The lines in (b) are the fittings to the R_dc behavior with m as indicated with arrows. The lines in (a) are the fittings to $C = C_1 \exp[qV_{oc}/m_C k_B T]$ with $m_C = 3.7$, and $C_1 = 5.0 \times 10^{-11}$ F·cm$^{-2}$ and $C_1 = 8.2 \times 10^{-11}$ F·cm$^{-2}$ for IS and LIMIS, respectively. (c) Low frequency limits of the voltage and current responsivities for different light intensities (see **Figure S12**) where the solid lines belong to fitting to $\Psi_V \propto P_{in}^{-1}$ and $\Psi_J \propto P_{in}^{0}$ and the dashed one to $\Psi_J \propto P_{in}^{-1/3}$

## S1.7. Transient photovoltage (TPV) measurements

*Setup description:* A Cree XP-E LED is used for white light bias. Driving the LED current with a Keithey 2400 and measuring the light intensity with a highly linear photodiode (Vishay BPW21R) allows to reproducibly adjust the light intensity with an error below 0.5% over a range of $10^{-5}$ to 1 suns. A small perturbation is induced with a 405 nm laser diode driven by a function generator from Agilent. The intensity of the short (50 ns) laser pulse is adjusted to keep the voltage perturbation below 10 mV, typically at 5 mV. After the pulse, the voltage decays back to its steady state value in a single exponential decay.[1] The characteristic decay time is determined from a linear fit to a logarithmic plot of the voltage transient and returns the small perturbation charge carrier lifetime.

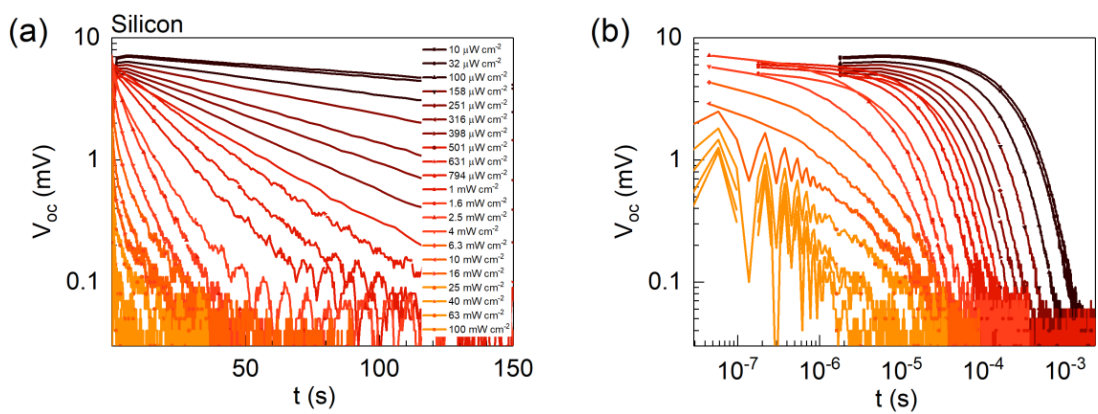

**Figure S15.** Experimental curves from the Zahner reference **silicon solar cell** (SiSC) processed in the transient photovoltage (TPV) setup at different illumination intensities, as indicated, in (a) semi-log and (b) log-log scales. In (b) is evident how at higher intensities the decays follow power laws rather than exponentials.

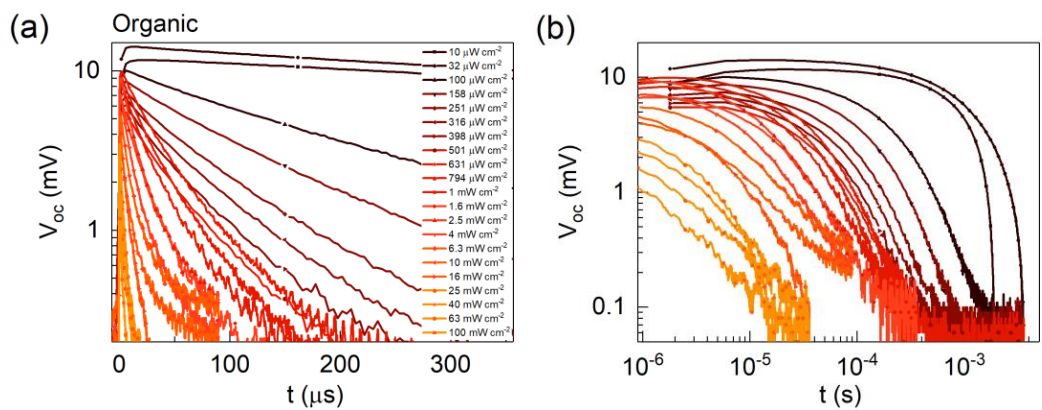

**Figure S16.** Experimental curves from the **organic solar cell** (OrgSC) processed in the TPV setup in (a) at different illumination intensities, as indicated, in (a) semi-log and (b) log-log scales. In (b) is evident how at higher intensities the decays follow power laws rather than exponentials.

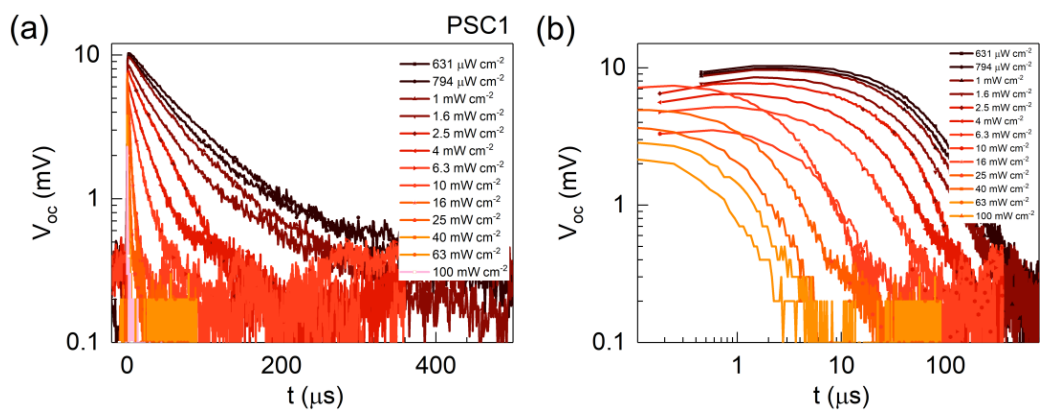

**Figure S17.** Experimental curves from the $SnO_2$/PMMA(PCBPM)/$Cs_{0.05}MA_{0.1}FA_{0.85}Pb(I_{0.85}Br_{0.15})_3$/PDCBT/$WO_x$/Au **perovskite solar cell** (PSC-1) processed in the TPV setup in (a) at different illumination intensities, as indicated, in (a) semi-log and (b) log-log scales. In (b) is evident how at higher intensities the decays follow power laws rather than exponentials.

## S1.8. Experimental J-V curves at different light intensities

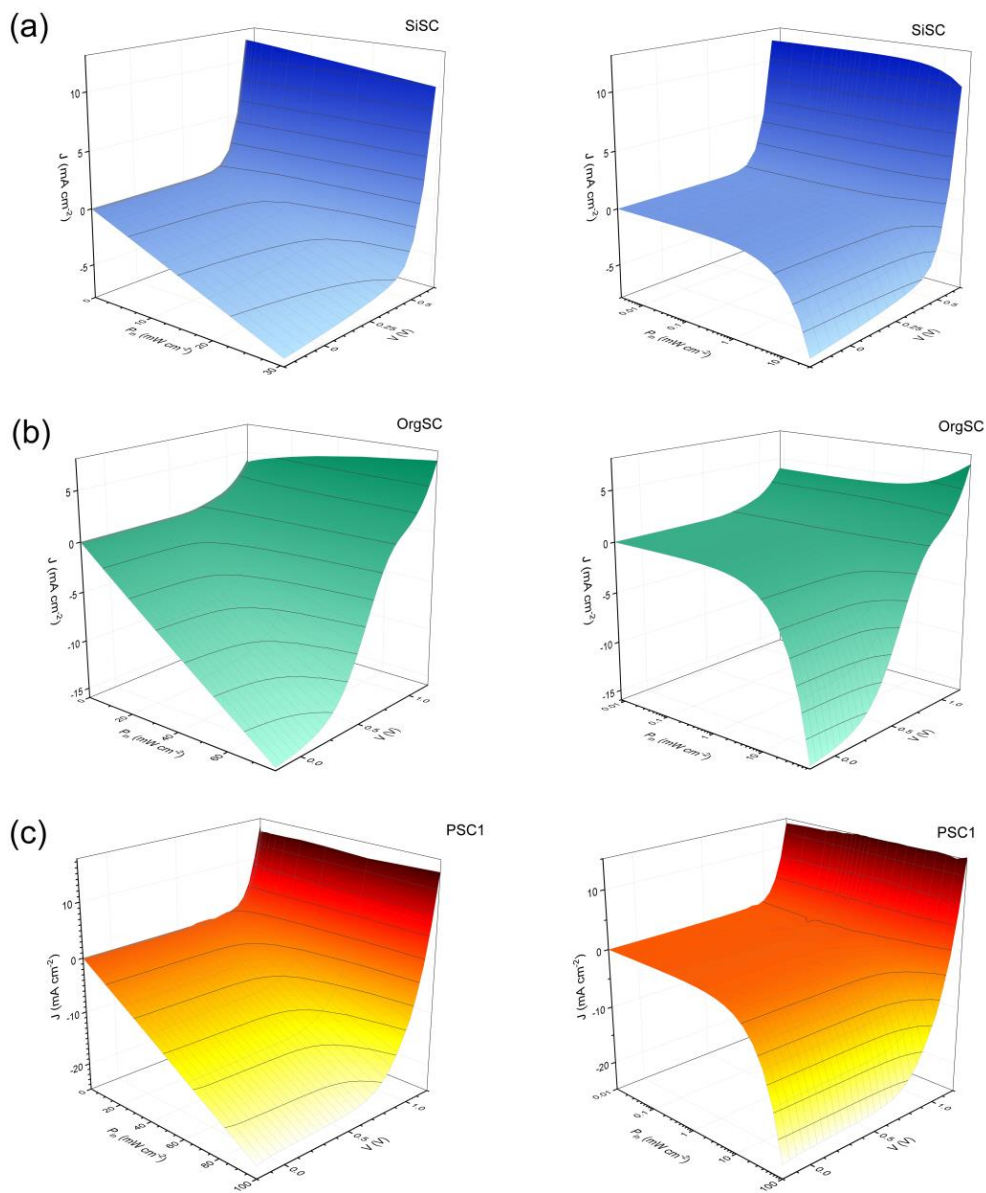

**Figure S18.** Experimental $J - V$ curves at different illumination intensities measured with the TPV setup in linear (left) and with log-scaled (right) illumination intensities from (a) SiSC, (b) OrgSC and (c) PSC1.

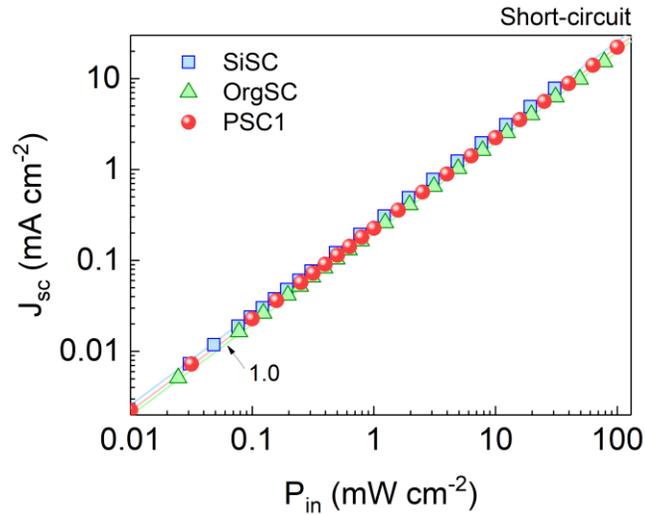

**Figure S19.** Experimental short-circuit current density as a function of the illumination intensity from the experimental data in **Figure S18** of the studied samples, as indicated. The arrow points the slope of the allometric fittings: $J_{sc} \propto P_{in}^{1}$.

# S2. Theory

### S.2.1. Numerical deduction of the *dc* resistances at OC for IS and LIMIS from the experimental *J-V* curves at different light intensities.

*Table S3:* Wolfram Mathematica code for the calculation of IS and LIMIS *dc* resistances at OC from experimental data in **Figure S18.**

```
SetDirectory["C:\\Users\\oalmora\\Dropbox\\2019_LIMIS"]; <<NumericalCalculus`

(*Silicon Solar cell (SiSC)*)
(*Importing data*)
Sidata=ToExpression[Import["Si.txt","Table"]];
Ps=Sidata[[1]] (*{y1,y2,y3...}*); Length[%];    Vs=Sidata[[2;;,1]] (*{x1,x2,x3...}*); Length[%];
Js=Sidata[[2;;,2;;]] ; Length[%];
Sixyzdata=Flatten[Table[{Vs[[j]],Ps[[i]],Js[[j,i]]},{j,Length[Vs]},{i,Length[Ps]}],1];
(*building function by interpolatind data*)
Jvp=Interpolation[Sixyzdata]
(*Plot3D[Jvp[x.y],{x,-0.2,0.8},{y,0,31}]*)
(*Finding the current roots for a given Voc*)
p0[Voc_]:=P/.FindRoot[Jvp[Voc,P],{P,1}]; p0[0.2]; p0[0.53](*control Pin values*);
(*Building function of voc(Pin) interpolating (roots, Voc)*)
Vocmin=0.2; Vocmax=0.52; dVoc=0.001;
VocP=Interpolation[Table[{p0[Voc],Voc},{Voc,Vocmin,Vocmax, dVoc/10}]]
LogLogPlot[VocP[P],{P,0.01,10},PlotRange->All,AxesLabel->{"Pin","Voc"}]
VocP[0.3](*control Voc value*);
(*defining derivative functions*)
dIS[Voc_]:=ND[Jvp[V,p0[Voc]],V,Voc]; dIMPS[Voc_]:=ND[Jvp[Voc,p],p,p0[Voc]];
```

```
dIMVS[Voc_]:=ND[VocP[p],p,p0[Voc]]
(*exporting the resistances from IS and LIMIS*)
RisSi=Table[1/dIS[Voc],{Voc,Vocmin,Vocmax, dVoc}] ; Export["R_IS_Si.txt",RisSi,"Table"]
RlimisSi=Table[dIMVS[Voc]/-dIMPS[Voc],{Voc,Vocmin,Vocmax,        dVoc}]          ;
Export["R_LIMIS_Si.txt",RlimisSi,"Table"]
VocList=Table[i,{i,Vocmin,Vocmax, dVoc}]; Export["VocList_Si.txt",VocList,"Table"]
PinList=Table[{p0[Voc],Voc},{Voc,Vocmin,Vocmax,                 dVoc}];
Export["PinList_Si.txt",PinList,"Table"]
ListLogPlot[{RisSi,RlimisSi}, PlotStyle->{Blue,RGBColor[0.74,0.,0.92]}]
```

(***Organic Solar cell (OrgSC)***)
```
(*Importing data*)
Orgdata=ToExpression[Import["Org.txt","Table"]];
Ps2=Orgdata[[1]]    (*{y1,y2,y3...}*);   Length[%];   Vs2=Orgdata[[2;;,1]]    (*{x1,x2,x3...}*);
Length[%]; Js2=Orgdata[[2;;,2;;]] ; Length[%];
Orgxyzdata=Flatten[Table[{Vs2[[j]],Ps2[[i]],Js2[[j,i]]},{j,Length[Vs2]},{i,Length[Ps2]}],1];
(*building function by interpolatind data*)
Jvp2=Interpolation[Orgxyzdata]
(*Plot3D[Jvp2[V,P],{V,-0.2,0.63},{P,0,75},AxesLabel->{"V","Pin","J"}]*)
(*Finding the current roots for a given Voc*)
p02[Voc_]:=P/.FindRoot[Jvp2[Voc,P],{P,80}]
(*Building function of voc(Pin) interpolating (roots, Voc)*)
Vocmin2=0.4; Vocmax2=0.8;dVoc2=0.01;
VocP2=Interpolation[Table[{p02[Voc],Voc},{Voc,Vocmin2,Vocmax2, dVoc2/10}]]
LogLogPlot[VocP2[P],{P,0.01,80},PlotRange->All,AxesLabel->{"Pin","Voc"}]
(*defining derivative functions*)
dIS2[Voc_]:=ND[Jvp2[V,p0[Voc]],V,Voc]; dIMPS2[Voc_]:=ND[Jvp2[Voc,p],p,p02[Voc]];
dIMVS2[Voc_]:=ND[VocP2[p],p,p02[Voc]]
(*exporting the resistances from IS and LIMIS*)
RisOrg=Table[1/dIS2[Voc],{Voc,Vocmin2,Vocmax2,               dVoc2}]                 ;
Export["R_IS_Org.txt",RisOrg,"Table"]
RlimisOrg=Table[dIMVS2[Voc]/-dIMPS2[Voc],{Voc,Vocmin2,Vocmax2, dVoc2}] ;
Export["R_LIMIS_Org.txt",RlimisOrg,"Table"]
VocList2=Table[i,{i,Vocmin2,Vocmax2, dVoc2}];
Export["VocList_Org.txt",VocList2,"Table"]
PinList2=Table[{p02[Voc],Voc},{Voc,Vocmin2,Vocmax2, dVoc2}];
Export["PinList_Org.txt",PinList2,"Table"]
ListLogPlot[{RisOrg,RlimisOrg}, PlotStyle->{Blue,RGBColor[0.74,0.,0.92]}]
```

(***Perovskite Solar cell 1 (PSC1)***)
```
(*Importing data*)
PSC1data=ToExpression[Import["PSC1.txt","Table"]];    Ps3=PSC1data[[1]]    (*{y1,y2,y3...}*);
Length[%];  Vs3=PSC1data[[2;;,1]]    (*{x1,x2,x3...}*);  Length[%];  Js3=PSC1data[[2;;,2;;]]   ;
Length[%];
PSC1xyzdata=Flatten[Table[{Vs3[[j]],Ps3[[i]],Js3[[j,i]]},{j,Length[Vs3]},{i,Length[Ps3]}],1];
(*building function by interpolating data*)
Jvp3=Interpolation[PSC1xyzdata]
Plot3D[Jvp3[V,P],{V,-0.2,1.25},[2],AxesLabel->{"V","Pin","J"}]
(*Finding the current roots for a given Voc*)
p03[Voc_]:=P/.FindRoot[Jvp3[Voc,P],{P,2}];          p03[0.80]; p03[1.1] (*control Pin values*);
(*Building function of voc(Pin) interpolating (roots, Voc)*)
Vocmin3=0.88; Vocmax3=1.10; dVoc3=0.01;
VocP3=Interpolation[Table[{p03[Voc],Voc},{Voc,Vocmin3,Vocmax3, dVoc3}]];
LogLogPlot[VocP3[P],{P,2.5,90},PlotRange->All,AxesLabel->{"Pin","Voc"}]
VocP3[3](*control Voc value*);
(*defining derivative functions*)
```

```
dIS3[Voc_]:=ND[Jvp3[V,p03[Voc]],V,Voc];        dIMPS3[Voc_]:=ND[Jvp3[Voc,p],p,p03[Voc]];
dIMVS3[Voc_]:=ND[VocP3[p],p,p03[Voc]]
(*exporting the resistances from IS and LIMIS*)
RisPSC1=Table[1/dIS3[Voc],{Voc,Vocmin3,Vocmax3, dVoc3}] ;
Export["R_IS_PSC1.txt",RisPSC1,"Table"]
RlimisPSC1=Table[dIMVS3[Voc]/-dIMPS3[Voc],{Voc,Vocmin3,Vocmax3, dVoc3}] ;
Export["R_LIMIS_PSC1.txt",RlimisPSC1,"Table"]
VocList3=Table[i,{i,Vocmin3,Vocmax3, dVoc3}];
Export["VocList_PSC1.txt",VocList3,"Table"]
PinList3=Table[{p03[Voc],Voc},{Voc,Vocmin3,Vocmax3, dVoc3}];
Export["PinList_PSC1.txt",PinList3,"Table"]
ListLogPlot[{RisPSC1,RlimisPSC1}, PlotStyle->{Blue,RGBColor[0.74,0.,0.92]}]
```

## S.2.2. Analytical deduction of the *dc* resistance from the light-bias corrected empirical Shockley equation.

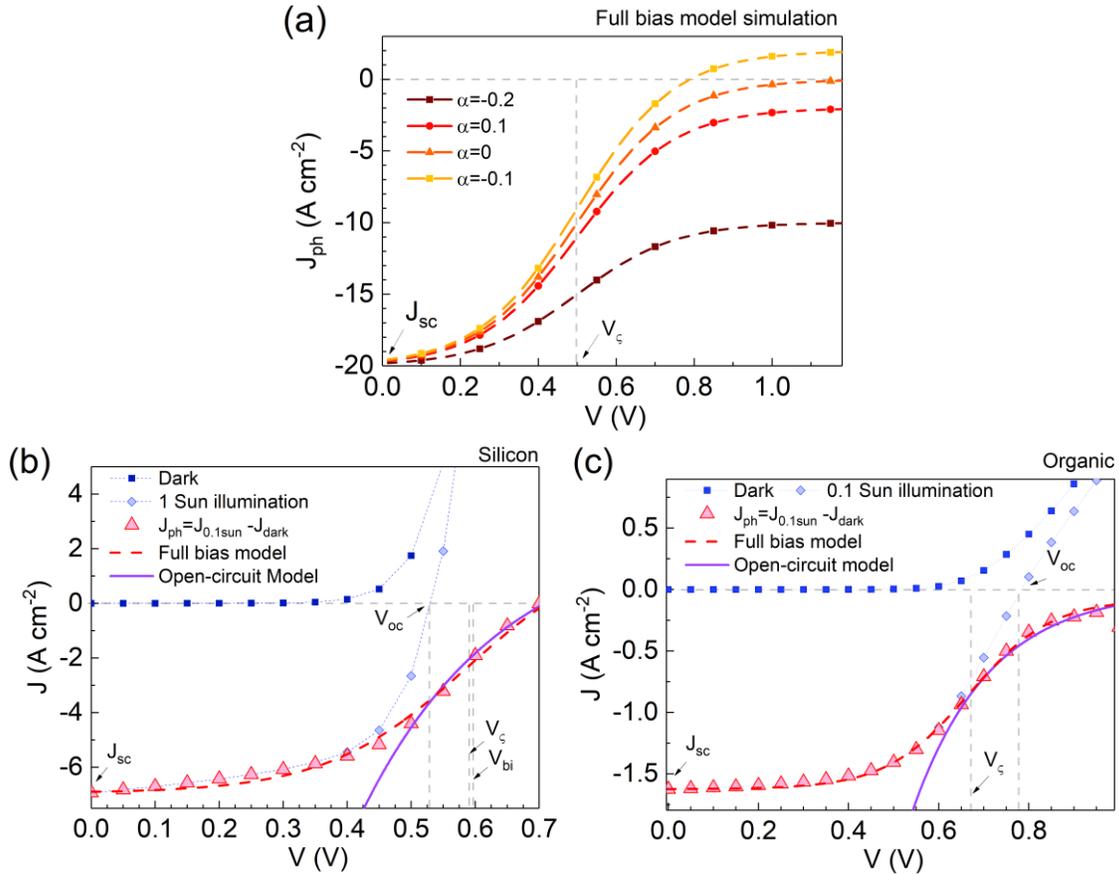

**Figure S20.** Bias dependent photocurrent: (a) full bias model simulation for different collection fractions and experimental curves and fittings for (b) SiSC and (c) OrgSC. Experimental data taken from **Figure S18**. Full bias model refers to equation (18) in the main manuscript $J_{ph} = J_{sc}\left(\varsigma + (\varsigma - 1)/(1 + \exp[q(V - V_\varsigma)/m_\Psi k_B T])\right)$ and open circuit model implies equation (20) in the main manuscript $J_{ph} \cong J_{sc}\exp[-q(V - V_\varsigma)/m_\Psi k_B T]$.

The empirical Shockley equation as typically used, neglecting parasitic resistance effects, is

$$J \cong J_s\left(\exp\left[\frac{qV}{m\,k_BT}\right] - 1\right) - P_{in}\Psi_{sc} \tag{S1}$$

where $q$ is the elementary electric charge, $k_BT$ the thermal energy, $J_s$ the saturation current, $m$ the ideality factor, $\Psi_J$ is the photo-current responsivity, and $P_{in}$ is the light intensity in units of power density. At open-circuit abode $5\,k_BT$ the above expression can be approached as

$$P_{in} \cong \frac{J_s}{\Psi_{sc}}\exp\left[\frac{qV_{oc}}{m\,k_BT}\right] \tag{S2}$$

Now, we consider the corrected empirical Shockley equation around the open circuit regime as:

$$J \cong J_s\left(\exp\left[\frac{qV}{m\,k_BT}\right] - 1\right) - P_{in}\Psi_{oc}\exp\left[-\frac{qV}{m_\Psi k_BT}\right] \tag{S3}$$

The corresponding open-circuit voltage for forward biases abode $5\,k_BT$ can be approached as

$$V_{oc} \cong \frac{m_\Psi m}{m_\Psi + m}\left(\frac{k_BT}{q}\ln\left[\frac{P_{in}\Psi_{oc}}{J_s}\right]\right) \tag{S4}$$

Subsequently, the derivative $\partial J/\partial V$ is obtained for (S3) as

$$\frac{\partial J}{\partial V} = \frac{qJ_s}{m\,k_BT}\exp\left[\frac{qV}{m\,k_BT}\right] + \frac{qP_{in}\Psi_{oc}}{m_\Psi\,k_BT}\exp\left[-\frac{qV}{m_\Psi k_BT}\right] \tag{S5}$$

$$\frac{\partial J}{\partial V} = \frac{qJ_s}{m\,k_BT}\exp\left[\frac{qV}{m\,k_BT}\right]\left(1 + \frac{m}{m_\Psi}\frac{P_{in}\Psi_{oc}}{J_s}\exp\left[-\frac{q}{k_BT}V\left(\frac{m_\Psi + m}{m_\Psi m}\right)\right]\right)$$

$$\frac{\partial J}{\partial V} = \frac{qJ_s}{m\,k_BT}\exp\left[\frac{qV}{m\,k_BT}\right]\left(1 + \frac{m}{m_\Psi}r_\Psi(V)\right)$$

with the photo-resistance factor

$$r_\Psi = \frac{P_{in}\Psi_{oc}}{J_s}\exp\left[-\frac{q}{k_BT}V\left(\frac{m_\Psi + m}{m_\Psi m}\right)\right] \tag{S6.a}$$

Now, assuming $\Psi_{oc} \approx \Psi_{sc}\exp[qV_\varsigma/m_\Psi k_BT]$, with the collection threshold voltage $V_\varsigma$ close to the built-in voltage, then

$$r_\Psi = \exp\left[-\frac{q}{k_BT}\left(\frac{m_\Psi + m}{m_\Psi m}\right)(V - V_\Psi)\right] \tag{S6.b}$$

With

$$V_\Psi = \frac{k_BT}{q}\ln\left[\frac{J_{sc}}{J_s}\right]\left(\frac{m_\Psi m}{m_\Psi + m}\right) + V_\varsigma\left(\frac{1}{1 + \frac{m_\Psi}{m}}\right) \tag{S6.c}$$

Note that $r_\Psi \to 0$ as $V$ increases but it is $r_\Psi > 1$ while $V < V_\Psi$. Subsequently, we can calculate the *dc* resistance from the IS concept as $1/(\partial J/\partial V)$:

$$R_{dc} \cong R_{th}\exp\left[-\frac{qV}{m\,k_BT}\right]\left(\frac{1}{1 + \frac{m}{m_\Psi}r_\Psi(V)}\right) \tag{S7}$$

where $R_{th} = m\,k_BT/J_s q$ is the thermal recombination resistance.

Following the concept of IMPS, the derivative $\partial J/\partial P_{in}$ is obtained for (S3) as

$$\Psi_{J,dc} = \Psi_{oc} \exp\left[-\frac{qV}{m_\Psi k_B T}\right] \tag{S8}$$

Definition (S8) can explain the decrease of $\Psi_J$ as $P_{in}$ increase at OC without an explicit connection.

Moreover, using the concept of IMVS, the derivative $\partial V_{oc}/\partial P_{in}$ is obtained for (S4) as

$$\Psi_{V,dc} \cong \frac{m_\Psi m}{m_\Psi + m} \frac{k_B T}{q\, P_{in}} \tag{S9}$$

Definition (S9) keeps the experimental trend of $\Psi_V \propto P_{in}^{-1}$.

Finally, we can use (S9) and (S8) to obtain the *dc* resistance from the LIMIS concept as

$$R_{\Psi,dc} \cong \frac{m_\Psi m}{m_\Psi + m} \frac{k_B T}{q\, P_{in} \Psi_{oc}} \exp\left[\frac{qV}{m_\Psi k_B T}\right] \tag{S10}$$

Here, we can substitute $P_{in}$ from his corrected definition at OC in (S3), so

$$P_{in} \cong \frac{J_s}{\Psi_{oc}} \exp\left[\frac{qV}{k_B T}\left(\frac{1}{m} + \frac{1}{m_\Psi}\right)\right] \tag{S11}$$

Subsequently, by substituting (S11) in (S10) we can rewrite the latter as

$$R_{\Psi,dc} \cong \frac{m_\Psi}{m_\Psi + m} R_{th} \exp\left[-\frac{qV}{m k_B T}\right] \tag{S12}$$

# Author Information

Corresponding Author

*E-mail: osbel.almora@fau.de, almora@uji.es,

# ORCID

Osbel Almora: 0000-0002-2523-0203

Germà Garcia-Belmonte: 0000-0002-0172-6175

Christoph J. Brabec: 0000-0002-9440-0253